\documentclass[preprint]{iucr}

\usepackage{graphicx}
\usepackage{xspace}
\usepackage{xcolor}
\usepackage{amsmath, amsthm, amssymb, amsfonts}

\usepackage{afterpage}
\usepackage{threeparttable}
\usepackage{siunitx}
\usepackage{longtable}
\usepackage{rotating}
\usepackage{booktabs}
\usepackage{comment}
\usepackage[normalem]{ulem}

\newcommand{\ins}[1]{\textcolor{brown}{[ins:#1]}}

\newcommand{\dd}{\mathrm{d}}

\newcommand{\nba}[1]{}

\definecolor{dgreen}{HTML}{008000}

\newcommand{\be}{\begin{enumerate}}
\newcommand{\ee}{\end{enumerate}}
\newcounter{saveenumi}

\newcommand{\qmax}{\ensuremath{Q_{\mathrm{max}}}\xspace}
\newcommand{\qmin}{\ensuremath{Q_{\mathrm{min}}}\xspace}

\newcommand{\pdfgui}{\textsc{PDFgui}\xspace}

\newcommand{\pdfgetxthree}{\textsc{PDFgetX3}\xspace}

\newcommand{\pdfgetn}{\textsc{PDFgetN}\xspace}
\newcommand{\fittwod}{\textsc{Fit2D}\xspace}

\newcommand{\xpdf}{\textsc{xPDFsuite}\xspace}
\newcommand{\cmi}{\textsc{DiffPy-CMI}\xspace}

\newcommand{\rw}{$R_w$\xspace}
\newcommand{\sto}{stoichiometry\xspace}
\newcommand{\sm}{structure-mining\xspace}

\renewcommand{\sout}[1]{}
\renewcommand{\ins}[1]{#1}

\makeatletter
\newcommand{\floatcaption}{%
\ifx \@captype \@undefined \@latex@error {\noexpand \caption outside float}\@ehd \expandafter \@gobble \else \refstepcounter \@captype \expandafter \@firstofone \fi {\@dblarg {\@caption \@captype }}%
}%
\makeatother

     \journalcode{A}              

\begin{document}                  


\title{Structure-mining: screening structure models by automated fitting to the atomic pair distribution function over large numbers of models}
\shorttitle{structure-mining}



\author[a]{Long}{Yang}
\author[b]{Pavol}{Juh\'{a}s}
\author[c]{Maxwell~W.}{Terban}
\author[d]{Matthew~G.}{Tucker}
\cauthor[a,e]{Simon~J.~L.}{Billinge}{sb2896@columbia.edu}


\aff[a]{Department of Applied Physics and Applied Mathematics, Columbia University, \city{New York}, NY 10027, \country{USA}}
\aff[b]{Computational Science Initiative, Brookhaven National Laboratory, \city{Upton}, NY 11973, \country{USA}}
\aff[c]{Max Planck Institute for Solid State Research, Heisenbergstrasse 1, 70569 \city{Stuttgart}, \country{Germany}}
\aff[d]{Neutron Scattering Division, Oak Ridge National Laboratory, \city{Oak Ridge}, TN 37830, \country{USA}}
\aff[e]{Condensed Matter Physics and Materials Science Department, Brookhaven National Laboratory, \city{Upton}, NY 11973, \country{USA}}









\maketitle                        


\begin{abstract}
A new approach is presented to obtain candidate structures from atomic pair distribution function (PDF) data in a highly automated way. It fetches, from web-based structural databases, all the structures meeting the experimenter's search criteria and performs structure refinements on them without human intervention. It supports both x-ray and neutron PDFs. Tests on various material systems show the effectiveness and robustness of the algorithm in finding the correct atomic crystal structure.  It works on crystalline and nanocrystalline materials including complex oxide nanoparticles and nanowires, low-symmetry and locally distorted structures, and complicated doped and magnetic materials. This approach could greatly reduce the traditional structure searching work and enable the possibility of high-throughput real-time auto analysis PDF experiments in the future.

\end{abstract}

\section{Introduction}

The development of science and technology is built on advanced materials, and new materials lie at the heart of technological solutions to major global problems such as sustainable energy~\cite{MoskowitzAdvancedMaterialsRevolution2014}. However, the discovery of new materials still needs a lot of labor and time.
The idea behind materials genomics~\cite{WhiteMaterialsGenomeInitiative2012} is to develop collaborations between materials scientists, computer scientists, and applied mathematicians to accelerate the development of new materials through the use of advanced computation such as artificial intelligence (AI), for example, by predicting undiscovered materials with interesting properties~\cite{JainCommentaryMaterialsProject2013a,Simonmaterialsgenomeaction2015,Curtarolohighthroughputhighwaycomputational2013}.

The study of material structure plays a key role in the development of novel materials. Structure solution of well ordered crystals is largely a solved problem, but for real materials, which may be defective or nanostructured, being studied under real conditions, for example in high-throughput \textit{in situ} and \textit{operando} diffraction experiments such as \textit{in situ} synthesis~\cite{CravillonFastNucleationGrowth2011a,JensenRevealingMechanismsSnO22012a,FriscicRealtimesitumonitoring2013,SahaSituTotalXRay2014d,Shoemakersitustudiesplatform2014,KatsenissituXraydiffraction2015,OldsCombinatorialappraisaltransition2017a,TerbanEarlystagestructural2018a}, determining structure can be a major challenge that could itself benefit from a genomics style approach.  Here we explore a data-mining methodology for the determination of inorganic \sout{materials}\ins{material} structures. The approach can rapidly screen large numbers of structures in a manner that is well matched to the kinds of high-throughput experiments being envisaged in the materials genomics arena.

A number of structural databases are available for inorganic materials containing structures solved from experimental data such as the Inorganic Crystal Structure Database (ICSD)~\cite{Bergerhoffinorganiccrystalstructure1983a,BelskyNewdevelopmentsInorganic2002}, the American Mineralogist Crystal Structure Database (AMCSD)~\cite{DownsAmericanMineralogistcrystal2003}, the Crystal Structure Database for Minerals (MINCRYST)~\cite{ChichagovMINCRYSTcrystallographicdatabase2001}, and the Crystallography Open Database (COD)~\cite{GrazulisCrystallographyOpenDatabase2009a}.
More recently, databases of theoretically predicted structures have begun to become available, such as the Materials Project Database (MPD)~\cite{JainCommentaryMaterialsProject2013a}, the Automatic Flow Library (AFLOWLIB)~\cite{CurtaroloAFLOWLIBORGdistributed2012}, and the Open Quantum Materials Database (OQMD)~\cite{SaalMaterialsDesignDiscovery2013,KirklinOpenQuantumMaterials2015}.
Structural databases such as the International Centre for Diffraction Data~\cite{icdd2019}, have for some time been used for phase identification purposes.
In phase identification studies no model fitting is carried out, but phases are identified in a powder \ins{diffraction} pattern by matching sets of the strongest Bragg peaks from the database structures to peaks in the  measured diffractogram~\cite{HanawaltChemicalAnalysisXRay1938b,Marquartsearchmatchsystem1979,GilmoreHighthroughputpowderdiffraction2004}. Our goal is not just phase identification, but \sout{the high-throughput automated refinement of structural models fit to measured diffraction data.  In our implementation we fit measured atomic pair distribution function (PDF) data, which has the additional benefit of allowing us to model on the fly nanostructured materials as well as crystalline materials.}\ins{to speed up the process of finding structural candidates to unknown atomic pair distribution function (PDF) signals.}

PDF analysis of x-ray and neutron powder diffraction datasets has been demonstrated to be an excellent tool for studying structures of many advanced materials, especially nanostructured materials~\cite{ZhangWaterdrivenstructuretransformation2003,NederStructurenanoparticlespowder2005b,MasadehQuantitativesizedependentstructure2007d,YoungApplicationspairdistribution2011a,beech;jacs14,terba;ic17,LavedaStructurepropertyinsights2018}, but also bulk crystalline materials~\cite{TobyOrderingTl2CaBa2Cu2O8Tl2Ba2CuO61989,BillingeDirectObservationLattice1996a,billi;cc04,Keencrystallographycorrelateddisorder2015a}.

The PDF gives the scaled probability of finding two atoms in a material a
distance $r$ apart and is related to the density of atom pairs in the
material. It does not presume periodicity so goes well beyond just well ordered crystals~\cite{egami;b;utbp12,billi;b;itoch18}.
The experimental PDF, denoted $G(r)$, is the \qmin and \qmax truncated Fourier transform of powder diffraction data:~\cite{farro;aca09}
\begin{equation}
\label{eq:FTofSQtoGr}
  G(r) = \frac{2}{\pi}
          \int_{\qmin}^{\qmax} Q[S(Q)-1]\sin(Qr) \: \dd Q,
\end{equation}
where $Q$ is the magnitude of the scattering momentum. The total scattering structure function, $S(Q)$, is extracted from the Bragg and diffuse components of x-ray, neutron, or
electron powder diffraction pattern.

$G(r)$ can be calculated from a given structure model~\cite{egami;b;utbp12} and once the experimental PDFs are determined they can be analyzed through modeling. The PDF modeling is performed by adjusting the parameters of the structure model, such as the lattice parameters, \sout{atom}\ins{atomic} positions, and \sout{anisotropic}atomic displacement parameters, to maximize the agreement between the calculated PDF from the structure model and the experimental PDF.

A number of PDF structure modeling programs are available for crystalline or nanocrystalline inorganic materials~\cite{crans;b;pdtap08}.
Small box modeling programs use a small number of crystallographic parameters with a periodic structure model~\cite{egami;b;utbp12}. Three widely used examples are \pdfgui~\cite{farro;jpcm07}, TOPAS~\cite{CoelhoTOPASTOPASAcademicoptimization2018}, and \cmi~\cite{juhas;aca15}, among others~\cite{PetkovFITcomputerprogram1990,ProffenPDFFITprogramfull1999c,GaginCombinedfittingsmall2014b}.
Big box modeling programs, which move large numbers of atoms to minimize the difference between the observed and calculated PDFs, usually implement the reverse Monte Carlo (RMC) method~\cite{McGreevyReverseMonteCarlo1988,McGreevyReverseMonteCarlo2001a},
such as RMCProfile~\cite{TuckerRMCProfilereverseMonte2007b}, DISCUS~\cite{ProffenDISCUSprogramdiffuse1997b,PageBuildingrefiningcomplete2011b}, and FullRMC~\cite{AounFullrmcrigidbody2016}.
Other modeling programs use a hybrid approach where a large number of atoms are in the box, but the program refines only a small number of parameters,
such as EPSR~\cite{SoperPartialstructurefactors2005b}.

Though powerful for understanding structure of complex materials, PDF modeling and structure refinement \sout{is}\ins{are} difficult and present a steep learning curve for new users.
There are two major challenges.
The first is that PDF structure refinement requires a satisfactory plausible starting model to achieve a successful result.
The second is that the refinement process is a non-linear regression that is highly non-convex and generally requires significant user inputs to guide it to the best fit whilst avoiding overfitting.
A more automated refinement program such as we propose here needs to address both issues.

Model selection traditionally requires significant chemical knowledge and experience, but can be quite challenging when unknown impurities or reaction products are present in the sample.
To address the problem of phase identification, automated search-match algorithms for identifying phases in powder diffraction patterns have been developed and are widely used~\cite{HanawaltChemicalAnalysisXRay1938b,Marquartsearchmatchsystem1979,GilmoreHighthroughputpowderdiffraction2004}. There are also programs for helping find candidate \sout{structure}\ins{structures} from structural databases~\cite{BarrSNAP1Dcomputerprogram2004,TobyCMPRpowderdiffraction2005,AltomareQUALXcomputerprogram2008,DegenHighScoresuite2014a,AltomareQUALX2qualitativephase2015a}.
These search-match programs only work for reciprocal space diffraction patterns, and in general do not allow for automated refinement of structures.
Some attempts have been made to couple Rietveld refinement programs to structural databases such as Full Profile Search Match~\cite{BoullayFastmicrostructurephase2014,LutterottiFullprofilesearchmatch2019}, though this is limited to refining structures from the COD database.  Alternatively, programs that use scripting such as TOPAS~\cite{CoelhoTOPASTOPASAcademicoptimization2018} have been used to automatically refine large numbers of candidate structures generated by symmetry-mode analysis from a given high-symmetry starting structure~\cite{LewisExhaustiveSymmetryApproach2016}.  Furthermore, a structure screening approach where large numbers of algorithmically generated small metal nanoparticle models were compared to PDF data was recently demonstrated~\cite{BanerjeeClusterminingapproachdetermining2020}.  This approach, called cluster-mining, was successful at obtaining significantly improved fits over standard approaches to nanoparticle PDF data from simple models with a small number of refinable parameters.  It also returned multiple plausible and well performing structures rather than just one best-fit structure, allowing the user to choose a model based on more information than just the PDF data.  We would like to combine these approaches (database searching, auto-refinement, and screening of large numbers of structures) to the modeling of PDF data in general.

Here we describe an approach we call \sm, to automate and manage structure model selection and PDF refinement. To make the whole procedure as high-throughout and automatic as possible, the required user inputs are kept to a minimum: simply the experimental PDF data and the searching criterion used to \ins{filter} which structures \ins{to fetch from the} databases.
When finished, the best-fit candidate structures \sout{that were pulled}from the data mine are returned to the experimenter for further detailed investigations.
Structure-mining currently supports both x-ray and neutron PDF datasets. This software enables high-throughput auto-refinement that may be used right after the PDF is obtained at a synchrotron x-ray or neutron beamline, unlike more traditional human intensive approaches that typically take a large amount of time and effort after the experiment is over. It is designed to lighten the PDF modeling work after an experiment, but could also, in principle, be used for modeling PDF datasets in quasi-real-time during the data acquisition at the beamline.

\section{Approach} \label{sec;approach}

Structure-mining first obtains a large number of candidate structures from open structural databases. It then computes the PDFs of these structures and carries out structure refinements to obtain the best agreement between calculated PDFs and the measured PDF under study.
The initial implementation uses two commonly utilized open structural databases: the Materials Project Database (MPD)~\cite{JainCommentaryMaterialsProject2013a} and the Crystallography Open Database (COD)~\cite{GrazulisCrystallographyOpenDatabase2009a}. The structures are \sout{pulled}\ins{fetched} directly from the databases using the RESTful API~\cite{OngPythonMaterialsGenomics2013,OngMaterialsApplicationProgramming2015}.
There are many rules that could be used for selecting candidate structures to try. In this initial implementation of \sm, we are using the following heuristics \ins{for filtering which structure models to fetch}: (1) \sout{Pulling}all the structures that have the same stoichiometry as prescribed by the experimenter, (2) \sout{Pulling}all the structures containing a prescribed list of elements, (3) \sout{Pulling}all the structures containing the prescribed list of elements plus a number of additional elements specified by a wild-card symbol, (4) \sout{Pulling}all the structures containing a subset of the prescribed elements plus other elements if a wild-card symbol is specified.   These heuristics go from more restrictive to less restrictive and may be selected as desired. The results on representative datasets are presented below.

After \sout{pulling}\ins{fetching} the structures\sout{from the database}, \sm builds a list of candidate structures and loads their cif files from the database into the \cmi~\cite{juhas;aca15} PDF structure refinement program.

\cmi works by first building a fit recipe which is the set of information needed to run a model refinement to PDF data, and then executing it.  The PDF fit recipe for each \sout{pulled}structure is generated automatically. \sout{The fits are carried out over the range of $1.5<r<20$~\AA\ on the Nyquist-Shannon sampling grid~\cite{farro;prb11} \ins{by default}.}\ins{By default, the fits are carried out over the range of $1.5<r<20$~\AA\ on the Nyquist-Shannon sampling grid~\cite{farro;prb11}, however, a different fit range may be specified by the user.}
The following phase related parameters are initialized and refined: a single scale factor uses initial value 1.0; lattice parameters are constrained according to the crystal systems using the initial lattice parameter values of \sout{pulled}the structures; isotropic atomic displacement parameter (ADP), $U_{iso}$, for each element atom of the \sout{pulled}structure is applied with initial value 0.005~\AA$^2$; spherical particle diameter (SPD) parameter can be used if the PDF data are from nano-sized objects, by having the experimenter specify an initial value (in the unit of \AA).
The instrument resolution parameters, $Q_{damp}$ and $Q_{broad}$, which are the parameters that correct the PDF envelope function for the instrument resolution~\cite{ProffenPDFFITprogramfull1999c,farro;jpcm07}, are preferably obtained by measuring a standard calibration material in the same experimental setup geometry as the measured sample, and are fixed in the subsequent structure refinements of the measured sample PDF.
They are applied according to the following strategy. If the experimenter specifies $Q_{damp}$ and $Q_{broad}$ values, the experimenter's values are used and they are fixed during the structure refinement. If they are not specified by the experimenter, the program will make a best-effort attempt to allocate meaningful values.  This is done currently by storing a table of reasonable values by instruments. So far, we have established reasonable values for the XPD x-ray instrument and the NOMAD and NPDF neutron instruments. If the program cannot find reasonable values in its lookup table for a specified instrument, or if no instrument can be determined, standard global default values are selected. These are $Q_{damp} = 0.04$~\AA$^{-1}$ for rapid acquisition x-ray PDF (RAPDF) experiments~\cite{ChupasRapidacquisitionpairdistribution2003c} and 0.02~\AA$^{-1}$ for time-of-flight (TOF) neutron PDFs.  Similarly, $Q_{broad}= 0.01$~\AA$^{-1}$ and 0.02~\AA$^{-1}$ are the global defaults for RAPDF x-ray and TOF neutron measurements, respectively.  In all the cases where the user does not specify values for $Q_{damp}$ and $Q_{broad}$, these parameters are allowed to vary in the refinement process.

Different regression algorithms may be used to perform the structure refinement minimizing the fit residual, with the goodness-of-fit \rw, given by
\begin{equation}
\label{eq:GoodnessOfFit}
  R_w = \sqrt{   \frac{\sum_{i=1}^{n}[G_{obs}(r_i) - G_{calc}(r_i,P)]^2}  {\sum_{i=1}^{n} G_{obs}(r_i)^2}    },
\end{equation}
where $G_{obs}$ and $G_{calc}$ are the observed and calculated PDFs and $P$ is the set of parameters refined in the model.

Initially we use the widely applied damped least-squares method (Levenberg\--Marquardt algorithm)~\cite{Levenbergmethodsolutioncertain1944,MarquardtAlgorithmLeastSquaresEstimation1963}, which is deployed in the Python programming package Scipy~\cite{JonesSciPyOpensource2001}, to vary the adjustable parameters to achieve the best agreement between the calculated and measured PDFs, since none of the algorithms for nonlinear least-squares problems has been proved to be superior to this standard solution~\cite{young1993rietveld,FloudasEncyclopediaOptimization2001}, such as Gauss-Newton method~\cite{GaussTheoriamotuscorporum1809}, modified Marquardt method~\cite{FletcherMODIFIEDMARQUARDTSUBROUTINE1971}, and conjugate direction method~\cite{Powellefficientmethodfinding1964}. However, \cmi supports the use of different minimizers and the implementation with different optimizers will be tested in the future.
During the structure refinement different types of parameters have quite different characteristic behaviors. A systematic parameter turn-on sequence is important to achieve convergence because turning on unstable parameters too early can result in divergent fits or getting trapped at local false minima. To make the \sm highly automatic without any human intervention during the whole procedure, here we tested an automatic turn-on sequence that was suggested for conventional full-profile Rietveld refinement~\cite{young1993rietveld} as well as considering the difference between PDF and Rietveld refinement procedures.
The current \sm deploys the following parameter turn-on sequence. (1) Scale factor and lattice parameters are allowed to vary for up to 10 iterations, (2) isotropic ADPs are allowed to vary for up to 100 iterations, (3) if selected, the instrument resolution parameters, $Q_{damp}$ and $Q_{broad}$, are turned on for up to 100 iterations, and finally (4) if SPD is specified, it will then be turned on for up to 100 iterations. When the whole procedure is finished, if the refinement cannot converge, the refinement will stop, record the latest goodness-of-fit parameter $R_w$ value, and continue with the next \sout{pulled}structure. If the resulted $R_w > 1.0$ (unconverged fit), it would be marked as 1.0.

This process is repeated for every structure \sout{pulled}\ins{fetched} from databases.
When the program has looped over all the \sout{pulled}structures it returns a plot of best-fit goodness-of-fit parameters $R_w$ of each model. We call this plot the \sm map (see a representative plot later in Fig.~\ref{fig;BTO_BaTiO3_MPD_COD}).

The program also returns a detailed formatted table that is suitable for inserting into a manuscript summarizing the results of the \sm process. The experimenter can also select one or multiple structure model entries to save the corresponding results, figures of the data and the fit, the calculated and difference PDF data files, the initial and refined structures in cif format, and the values of initial and refined parameters in a formatted table.

\ins{Structure-mining will be made available on a cloud-based platform at https://pdfitc.org.}

\section{Testing the approach}
\subsection{testing methodology}

To test the method, we selected PDFs of \sout{five}\ins{seven} different materials \sout{, testing both x-ray and neutron PDFs}\ins{from x-ray and neutron total scattering data}, as listed in Table~\ref{tab;samples}.

\begin{table}
\caption{The experimental PDF datasets for testing the \sm approach.  The reference describing the experiments is given except for the Ti$_4$O$_7$ data which are unpublished.}
\label{tab;samples}

\begin{threeparttable}
\begin{tabular}{lll}
Composition & Scatterer & Beamline  \\
\hline

BaTiO$_3$\tnote{a} & x-ray & XPD  \\
Ti$_4$O$_7$ & x-ray & XPD \\
NaFeSi$_2$O$_6$\tnote{b} & x-ray & XPD \\
Ba$_{0.8}$K$_{0.2}$(Zn$_{0.85}$Mn$_{0.15}$)$_2$As$_2$\tnote{c} & neutron & NOMAD  \\
CuIr$_2$S$_4$\tnote{d} & x-ray & XPD \\
MnO\tnote{e} & neutron & NPDF  \\
V$_2$N+VN\tnote{f} & x-ray & XPD  \\
\hline

\end{tabular}
\begin{tablenotes}
\item [a] \cite{lombardi;cm19}.
\item [b] \cite{lewis;cec18}.
\item [c] \cite{frand;prb16}.
\item [d] \cite{BozinLocalorbitaldegeneracy2019c}.
\item [e] \cite{frand;aca15}.
\item [f] \cite{Urbankowski2Dmolybdenumvanadium2017d}.
\end{tablenotes}
\end{threeparttable}
\end{table}

The total scattering measurements were conducted at one synchrotron x-ray facility, the XPD beamline (28-ID-2) at the National Synchrotron Light Source II (NSLS-II), Brookhaven National Laboratory, and two neutron time-of-flight facilities, the NOMAD beamline (BL-1B)~\cite{NeuefeindNanoscaleOrderedMAterials2012a} at the Spallation Neutron Source (SNS) at Oak Ridge National Laboratory and the NPDF beamline~\cite{ProffenBuildinghighresolution2002} at the Manuel Lujan Jr. Neutron Scattering Center at Los Alamos Neutron Science Center (LANSCE), Los Alamos National Laboratory.  All of the datasets are from previously published work, indicated in the table, except for the Ti$_4$O$_7$, which is unpublished data.

For the XPD beamline the samples were loaded in 1~mm inner diameter polyimide capillaries mounted perpendicular to the beam and the x-ray datasets were collected at room temperature, except the vanadium nitride sample that was collected at 100~K~\cite{Urbankowski2Dmolybdenumvanadium2017d} and the CuIr$_2$S$_4$ sample at 500~K~\cite{BozinLocalorbitaldegeneracy2019c}, using the rapid acquisition PDF method (RAPDF)~\cite{ChupasRapidacquisitionpairdistribution2003c}. A large area 2D Perkin Elmer detector was mounted behind the samples. The collected data frames were summed, corrected for detector and polarization effects, and masked to remove outlier pixels before being integrated along arcs of constant $Q$, where  $Q=4\pi\sin\theta/\lambda$ is the magnitude of the momentum transfer on scattering, to produce 1D powder diffraction patterns using the \fittwod program~\cite{HammersleyFIT2Dmultipurposedata2016}. Standardized corrections and normalizations were applied to the data to obtain the total scattering structure function, $S(Q)$, which was Fourier transformed to obtain the PDF, using \pdfgetxthree~\cite{juhas;jac13} within \xpdf~\cite{yang;arxiv15}.
The incident x-ray wavelengths and the calibrated sample-to-detector distances are listed in the Appendix (Table~\ref{tab;sisamples}).

For the NOMAD and NPDF beamlines, the samples were loaded in vanadium cans. \sout{The NOMAD experiment was carried out}\ins{The Ba$_{0.8}$K$_{0.2}$(Zn$_{0.85}$Mn$_{0.15}$)$_2$As$_2$ data from the NOMAD beamline were collected} at room temperature~\cite{frand;prb16} and the data were reduced and transformed to the PDF using the automated data reduction scripts at the NOMAD beamline.
\sout{For the NPDF beamline, the data}\ins{The MnO data from the NPDF beamline} were collected at 15~K~\cite{frand;aca15} and the data were reduced and transformed to the PDF using the \pdfgetn program~\cite{PetersonPDFgetNuserfriendlyprogram2000}.

The full experimental details may be found in Refs.~\cite{lombardi;cm19,lewis;cec18,frand;prb16,BozinLocalorbitaldegeneracy2019c,frand;aca15,Urbankowski2Dmolybdenumvanadium2017d}. The maximum range of data used in the Fourier transformation, $Q_{max}$, was chosen to give the best trade-off between statistical noise and real-space resolution. The instrument resolution parameters, $Q_{damp}$ and $Q_{broad}$, which are relevant parameters for our structure-mining activity, were obtained by calibrating the experimental conditions in each case using a well crystallized standard sample.  The values are reproduced in the Appendix (Table~\ref{tab;sisamples}).

\subsection{Results}

\begin{figure}
\includegraphics[width=1\textwidth]{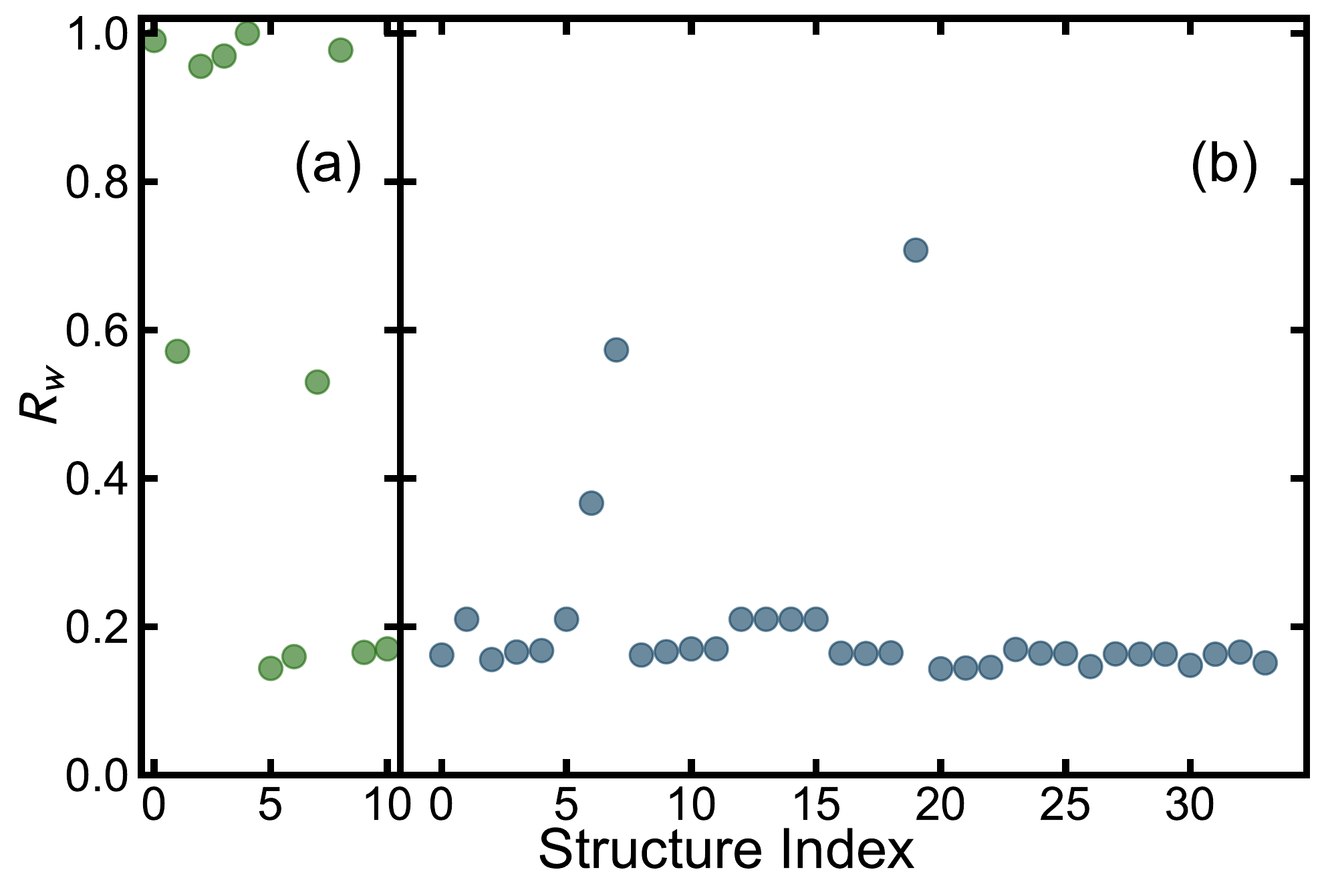}\\
\caption{\rw values for each of the structures \sout{pulled}\ins{fetched} from the databases for the BaTiO$_3$ nanoparticle x-ray data using heuristic-1, \sout{fetching}\ins{filtering for} all the structures with composition BaTiO$_3$ from (a) the MPD (green) and (b) the COD (blue). The \rw parameter represents the goodness-of-fit for each \sout{pulled}structure.}
\label{fig;BTO_BaTiO3_MPD_COD}
\end{figure}

\begin{figure}
\includegraphics[width=0.6\textwidth]{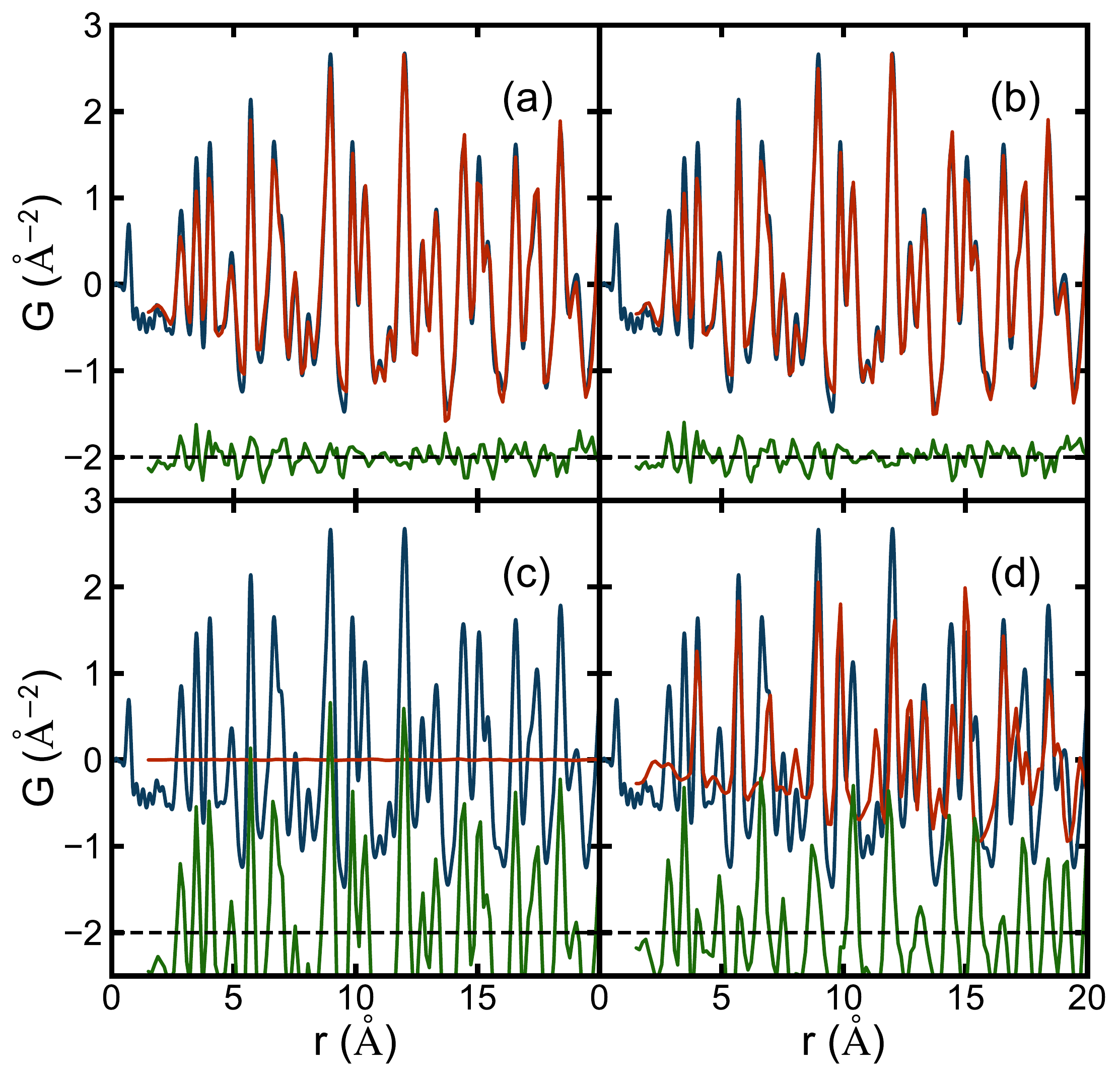}
\caption{PDFs from representative satisfactory and unsatisfactory structures from (a, c) the MPD  and (b, d) the COD. Blue curves are the measured PDF of BaTiO$_3$ nanoparticles.  Red curves are the calculated PDFs after retrieving from the databases using heuristic-1 and automatically fitting to the data (see Section~\ref{sec;approach} for details). Offsets below in green are the difference curves.
}
\label{fig;BTO_BaTiO3_MPD_COD_fits}
\end{figure}

We first apply this approach to the measured PDF from barium titanate (BTO) nanoparticles, BaTiO$_3$.  BTO is one of the best studied perovskite ferroelectric materials~\cite{FrazerSingleCrystalNeutronAnalysis1955,kwei;jpc93}.
Heuristic-1 is applied, fetching all structures that have the same composition as input BaTiO$_3$.
The \sm results from the MPD and COD are shown in Fig.~\ref{fig;BTO_BaTiO3_MPD_COD}(a) and (b), and Table~\ref{tab;BTO_BaTiO3_MPD} and Table ~\ref{tab;BTO_BaTiO3_COD}, respectively.
%
\begin{table*}
\small
\centering
\floatcaption{Structure-mining results for the BaTiO$_3$ nanoparticle x-ray data using heuristic-1 from the MPD. Here No. refers to the structure index (Fig.~\ref{fig;BTO_BaTiO3_MPD_COD}(a)), which is the order \sout{pulled}\ins{fetched} from the database, and s.g. represents the space group of the structure model. The initial isotropic atomic displacement parameter ($U_{iso}$) of all atoms in each structure is set to 0.005~\AA$^2$ to start the structure refinements. The $a$, $b$, and $c$ are the lattice parameters of the structure model. The subscript~$i$ indicates an initial value before refinement and the subscript~$r$ indicates a refined value. DB ID represents the database ID of the structure model. $Q_{max}=24.0$~\AA$^{-1}$, $Q_{damp}=0.037$~\AA$^{-1}$, and $Q_{broad}=0.017$~\AA$^{-1}$ were set and not varied in the refinements (see Section~\ref{sec;approach} for details).}
\label{tab;BTO_BaTiO3_MPD}
\scalebox{0.8}{
\begin{tabular}{cccccccccccccc}
\toprule
{No.} &        \rw  & s.g. &  Ba $U_{iso}$ &  Ti $U_{iso}$ &  O $U_{iso}$ &       $a_i$ &       $a_r$ &        $b_i$ &       $b_r$ &        $c_i$ &        $c_r$ &     DB ID \\
 &  & & {(\AA$^{2}$)} & {(\AA$^{2}$)} & {(\AA$^{2}$)}  & {(\AA)}  & {(\AA)}  & {(\AA)}  & {(\AA)}  & {(\AA)}  & {(\AA)} & \\

\midrule
          5 &  0.144  &        $Amm2$ &    0.0021 &   0.0070 &  0.0126 &  5.81 &  5.67 &   5.86 &  5.76 &   3.99 &   3.99 &     mp-5777 \\
          6 &  0.160  &        $P4mm$ &    0.0027 &   0.0074 &  0.0116 &  4.00 &  4.00 &      - &     - &   4.22 &   4.07 &     mp-5986 \\
          9 &  0.165  &         $R3m$ &    0.0027 &   0.0074 &  0.0123 &  5.75 &  5.66 &      - &     - &   7.11 &   7.05 &     mp-5020 \\
         10 &  0.170  &      $P4/mmm$ &    0.0026 &   0.0105 &  0.0174 &  4.03 &  4.00 &      - &     - &   4.04 &   4.07 &     mp-2998 \\
          7 &  0.530  &     $C222_1$ &    0.0047 &   0.0023 &  0.0373 &  5.84 &  5.69 &  10.02 &  9.84 &  14.14 &  13.98 &   mp-558125 \\
          1 &  0.571  &   $P6_3/mmc$ &    0.0070 &   0.0041 &  0.0468 &  5.79 &  5.69 &      - &     - &  14.10 &  13.97 &     mp-5933 \\
          2 &  0.956  &      $P4/mmm$ &    0.0172 &   0.0011 &  0.0884 &  4.11 &  4.16 &      - &     - &   5.04 &   4.73 &    mp-19990 \\
          3 &  0.969  &        $Amm2$ &    0.0003 &   0.0941 &  0.0090 &  5.31 &  5.26 &   5.33 &  5.44 &   8.88 &   8.80 &  mp-1076932 \\
          8 &  0.977  &        $Amm2$ &    0.0075 &   0.0006 &  0.0010 &  6.64 &  6.76 &   8.63 &  8.60 &   3.75 &   3.86 &   mp-644497 \\
          0 &  0.990  &        $Amm2$ &    0.0017 &   0.0031 &  0.0000 &  5.81 &  6.00 &   5.85 &  5.98 &   5.03 &   4.84 &   mp-995191 \\
          4 &  1.000  &       $Pm\overline{3}m$ &    0.0115 &  0.0104 &  0.0003 &  4.65 &  4.78 &      - &     - &      - &      - &   mp-504715 \\
\bottomrule
\end{tabular}
}
\end{table*}
%
\begin{table*}
\small
\centering
\floatcaption{Structure-mining results for the BaTiO$_3$ nanoparticle x-ray data using heuristic-1 from the COD. See the caption of Table~\ref{tab;BTO_BaTiO3_MPD} for an explanation of the entries.}
\label{tab;BTO_BaTiO3_COD}
\scalebox{0.8}{
\begin{tabular}{cccccccccccccc}
\toprule
{No.} &        \rw  & s.g. &  Ba $U_{iso}$ &  Ti $U_{iso}$ &  O $U_{iso}$ &       $a_i$ &       $a_r$ &        $b_i$ &       $b_r$ &        $c_i$ &        $c_r$ &     DB ID \\
 &  &  & {(\AA$^{2}$)} & {(\AA$^{2}$)} & {(\AA$^{2}$)}  & {(\AA)}  & {(\AA)}  & {(\AA)}  & {(\AA)}  & {(\AA)}  & {(\AA)} & \\

\midrule
         20 &  0.143  &        $Amm2$ &    0.0021 &    0.0080 &   0.0181 &  5.67 &  5.67 &  5.69 &  5.76 &   3.98 &   3.99 &  9014492 \\
         21 &  0.144  &        $Amm2$ &    0.0020 &    0.0087 &   0.0178 &  5.67 &  5.67 &  5.69 &  5.76 &   3.98 &   3.99 &  9014627 \\
         22 &  0.145  &        $Amm2$ &    0.0020 &    0.0091 &   0.0175 &  5.67 &  5.67 &  5.69 &  5.76 &   3.99 &   3.99 &  9014645 \\
         26 &  0.146  &        $Amm2$ &    0.0020 &    0.0094 &   0.0168 &  5.67 &  5.67 &  5.68 &  5.76 &   3.99 &   3.99 &  9014774 \\
         30 &  0.148  &        $Amm2$ &    0.0020 &    0.0100 &   0.0179 &  5.67 &  5.67 &  5.69 &  5.76 &   3.98 &   3.99 &  9016084 \\
         33 &  0.151  &        $Amm2$ &    0.0025 &    0.0083 &   0.0070 &  5.68 &  5.68 &  5.69 &  5.75 &   3.99 &   3.99 &  9016638 \\
          2 &  0.156  &        $P4mm$ &    0.0027 &    0.0064 &   0.0116 &  3.99 &  4.00 &     - &     - &   4.04 &   4.07 &  1513252 \\
          8 &  0.162  &        $P4mm$ &    0.0026 &    0.0086 &   0.0162 &  4.00 &  4.00 &     - &     - &   4.02 &   4.07 &  2100858 \\
          0 &  0.162  &        $P4mm$ &    0.0026 &    0.0086 &   0.0162 &  4.00 &  4.00 &     - &     - &   4.02 &   4.07 &  1507756 \\
         31 &  0.163  &         $R3m$ &    0.0027 &    0.0076 &   0.0163 &  5.65 &  5.66 &     - &     - &   6.96 &   7.05 &  9016152 \\
         29 &  0.163  &        $Amm2$ &    0.0028 &    0.0054 &   0.0029 &  5.62 &  5.63 &  5.64 &  5.70 &   4.01 &   4.06 &  9015715 \\
         28 &  0.163  &         $R3m$ &    0.0027 &    0.0077 &   0.0161 &  5.65 &  5.66 &     - &     - &   6.95 &   7.05 &  9015616 \\
         27 &  0.163  &         $R3m$ &    0.0027 &    0.0079 &   0.0160 &  5.65 &  5.66 &     - &     - &   6.95 &   7.05 &  9015236 \\
         25 &  0.164  &         $R3m$ &    0.0027 &    0.0082 &   0.0160 &  5.66 &  5.66 &     - &     - &   6.95 &   7.05 &  9014756 \\
         17 &  0.164  &         $R3m$ &    0.0027 &    0.0083 &   0.0158 &  5.66 &  5.66 &     - &     - &   6.96 &   7.05 &  9014179 \\
         24 &  0.164  &         $R3m$ &    0.0026 &    0.0084 &   0.0153 &  5.65 &  5.66 &     - &     - &   6.95 &   7.05 &  9014743 \\
         16 &  0.164  &         $R3m$ &    0.0026 &    0.0085 &   0.0157 &  5.66 &  5.66 &     - &     - &   6.95 &   7.05 &  9014074 \\
         18 &  0.165  &         $R3m$ &    0.0026 &    0.0087 &   0.0150 &  5.65 &  5.66 &     - &     - &   6.95 &   7.05 &  9014230 \\
         32 &  0.166  &         $R3m$ &    0.0026 &    0.0091 &   0.0149 &  5.65 &  5.66 &     - &     - &   6.96 &   7.05 &  9016624 \\
          3 &  0.166  &        $P4mm$ &    0.0026 &    0.0096 &   0.0151 &  3.99 &  4.00 &     - &     - &   4.03 &   4.07 &  1525437 \\
          9 &  0.166  &        $P4mm$ &    0.0026 &    0.0097 &   0.0158 &  4.00 &  4.00 &     - &     - &   4.02 &   4.07 &  2100859 \\
          4 &  0.168  &        $Pmm2$ &    0.0025 &    0.0095 &   0.0151 &  3.98 &  3.99 &  4.01 &  4.01 &   4.02 &   4.07 &  1540757 \\
         23 &  0.169  &        $P4mm$ &    0.0026 &    0.0103 &   0.0163 &  4.00 &  4.00 &     - &     - &   4.02 &   4.07 &  9014668 \\
         11 &  0.170  &      $P4/mmm$ &    0.0026 &    0.0105 &   0.0174 &  4.00 &  4.00 &     - &     - &   4.02 &   4.07 &  2100861 \\
         10 &  0.170  &      $P4/mmm$ &    0.0026 &    0.0105 &   0.0174 &  4.00 &  4.00 &     - &     - &   4.02 &   4.07 &  2100860 \\
         15 &  0.210  &       $Pm\overline{3}m$ &    0.0046 &     0.0132 &   0.0172 &  3.97 &  4.02 &     - &     - &      - &      - &  5910149 \\
          1 &  0.210  &       $Pm\overline{3}m$ &    0.0046 &     0.0132 &   0.0172 &  4.01 &  4.02 &     - &     - &      - &      - &  1507757 \\
         13 &  0.210  &       $Pm\overline{3}m$ &    0.0046 &     0.0132 &   0.0172 &  4.01 &  4.02 &     - &     - &      - &      - &  2100863 \\
         12 &  0.210  &       $Pm\overline{3}m$ &    0.0046 &     0.0132 &   0.0172 &  4.01 &  4.02 &     - &     - &      - &      - &  2100862 \\
          5 &  0.210  &       $Pm\overline{3}m$ &    0.0046 &     0.0132 &   0.0172 &  4.00 &  4.02 &     - &     - &      - &      - &  1542140 \\
         14 &  0.210  &       $Pm\overline{3}m$ &    0.0046 &     0.0132 &   0.0172 &  4.03 &  4.02 &     - &     - &      - &      - &  4124842 \\
          6 &  0.367  &       $Pm\overline{3}m$ &    0.0058 &     0.0126 &   0.0799 &  4.08 &  4.02 &     - &     - &      - &      - &  1542189 \\
          7 &  0.573  &    $P6_3/mmc$ &    0.0070 &    0.0041 &   0.0469 &  5.72 &  5.69 &     - &     - &  13.96 &  13.97 &  2009488 \\
         19 &  0.708  &        $P4mm$ &    0.0042 &    0.2490 &   0.0479 &  3.99 &  4.04 &     - &     - &   4.03 &   3.98 &  9014273 \\
\bottomrule
\end{tabular}
}
\end{table*}
The best-fit structures from each data mine were MPD structure No.~5~\cite{ShiraneNeutronDiffractionStudy1957a} and COD structure No.~20~\cite{kwei;jpc93} with $R_w = 0.144$ and 0.143, respectively.  The calculated and measured PDFs are shown in Fig.~\ref{fig;BTO_BaTiO3_MPD_COD_fits}(a) and (b), respectively. Unlike the traditional manual PDF structure refinement methodology, the \sm approach followed by the automated fitting resulted in satisfactory and reasonable fits without any human intervention.  These structures may be investigated in more detail by traditional manual fitting approaches.

Some structures retrieved from the mine also resulted in very poor fits, as shown in Fig.~\ref{fig;BTO_BaTiO3_MPD_COD_fits}(c) and (d), which
are the automatically determined fits of MPD structure No.~4~\cite{JainCommentaryMaterialsProject2013a} and COD structure No.~19~\cite{ShiraneNeutronDiffractionStudy1957a}, respectively.  We expect that this will be due to the fact that the structure \sout{pulled}\ins{fetched} from the database is different from that of our sample, and it is this automated screening of database structures to find the most plausible candidates that is the goal of \sm. However, we investigate this in more detail below.

The structure of this measured BaTiO$_3$ nanoparticle dataset has been carefully studied before~\cite{lombardi;cm19}.  In that work, it was reported that the structure of this nanoparticle sample was non-centrosymmetric \ins{at room temperature} and had one of the ferroelectric forms of the BaTiO$_3$ structures~\cite{kwei;jpc93}, among one of the distorted structures with space groups \textit{A}\textit{mm}2, \textit{P}4\textit{mm}, and \textit{R}3\textit{m}.   All these structures gave somewhat comparable fit to the data and it was \sout{not possible}\ins{difficult} to distinguish which among them was definitively the correct structure \ins{because of the Bragg peak broadening and the relatively weak x-ray scattering of oxygen sublattice}. Nearby centrosymmetric space groups also performed well based on \rw but could be ruled out by careful consideration of refined ADPs of Ti ions \ins{(we note that in table V of~\cite{kwei;jpc93} there is a typo where the s.g. \textit{P}4\textit{mm} structure is described as \textit{P}4/\textit{mmm}, which is the centrosymmetric parent, but in the body of the table the Ti ion is shown as displacing off the center of unit cell, breaking centrosymmetry).}

From the MPD result, as shown in Table~\ref{tab;BTO_BaTiO3_MPD}, it clearly reveals that the top three best-fit structures are exactly the non-centrosymmetric ferroelectric forms of BaTiO$_3$ structures with space groups \textit{A}\textit{mm}2, \textit{P}4\textit{mm}, and \textit{R}3\textit{m}. In addition, the closely similar centrosymmetric perovskite model with space group \textit{P}4/\textit{mmm} (No.~10, ranked 4)~\cite{SrilakshmiStructureCatalyticActivity2016} gives sightly worse but comparable \rw. The heuristic-1 has therefore found the correct candidate structure models from the MPD, as well as returning nearby structures for a more detailed manual comparison.

The COD contained many more candidate structures for this composition (Table~\ref{tab;BTO_BaTiO3_COD}). Again the \sm shows that the best three \ins{non-centrosymmetric} perovskite models \sout{with space groups \textit{A}\textit{mm}2, \textit{P}4\textit{mm}, and \textit{R}3\textit{m}}are found as expected, along with the similar general barium titanate perovskite models (with slightly worse \rw) with space groups \textit{P}4/\textit{mmm} and \textit{P}\textit{m}$\overline{3}$\textit{m}.

The COD result also returned a space group \textit{P}\textit{mm}2 structure (No.~4)~\cite{zengBGDX1991} with a reasonable fit ($R_w = 0.168$) which turns out to be a general perovskite structure having two half filled Ti ions at (0.5,0.5,0.509) and (0.5,0.5,0.491) sites,  similar to a doubled unit cell of the tetragonal barium titanate perovskite model with space group \textit{P}4\textit{mm}, albeit with a small orthorhombic distortion.  This illustrates the power of this \sm approach as it does a good job of finding all plausible structures in the database.  These can then be considered and ruled out by researchers based on other criteria.

There is also a hexagonal \sout{perovskite}structure (space group \textit{P}$6_3$/\textit{mmc}) in the databases for BaTiO$_3$, and this gives very poor fit to the BaTiO$_3$ nanoparticle data from both MPD (No.~1)~\cite{AkimotoRefinementhexagonalBaTiO31994} and COD (No.~7)~\cite{AkimotoRefinementhexagonalBaTiO31994}, showing that the approach is capable of finding true positive and true negative results.

The \sm gives the COD structure No.~19 (space group: \textit{P}4\textit{mm})~\cite{ShiraneNeutronDiffractionStudy1957a} a bad fit because the model is wrong, with Ti ion sitting at 1b (0.5, 0.5, 0.265) and O2 ion sitting at 2c (0.5, 0, 0.236), which is significantly offset from the correct position such that Ti ion is at or near the center of the unit cell.  We checked the reference for this database entry (COD ID: 9014273), and it turned out to be correct in the paper but a wrong entry in the database because the reference reported that Ti ion was at 1b (0.5, 0.5, 0.0.515) and O2 ion was at 2c (0.5, 0, 0.486)~\cite{ShiraneNeutronDiffractionStudy1957a}. This indicates that this \sm approach may actually help to find errors in the database, but at worst will not return incorrect structures as candidate models.

Interestingly, the mining operation did report one false negative.  It missed one of the plausible perovskite structure models in the MPD database, the cubic \sout{heterostructure}model with space group \textit{P}\textit{m}$\overline{3}$\textit{m} (MPD No.~4)~\cite{JainCommentaryMaterialsProject2013a}, which was correctly found in the COD database. The reason why this did not give a good refinement was that the starting lattice parameters taken from the database were much too large \ins{($a = 4.65$~\AA)} and the automated refinement could not converge to the correct minimum \ins{($a = 4.02$~\AA) due to the 55\% cell volume mismatch from the correct one}, resulting in a poor fit.
Although we refine the lattice parameter during the process, if the starting value is too far away from the correct one, it is possible that the refinement program will not be able to find the right solution in the parameter space and result in a poor fit and a false negative result. \sout{We could think of strategies for increasing the convergence in the future.  However, in}\ins{In} some respect it is a success of the program because we actually hope that incorrect models in the database will fit the data poorly, and if the value of the lattice parameter recorded in the database is far from being correct for the measured sample, in some sense this constitutes a bad model. Similar lattice parameter situations happen for MPD No.~0~\cite{XiaoCrystalstructuredense2008}, 2~\cite{Donohueeffectvarioussubstituents1958}, 3~\cite{XiaoCrystalstructuredense2008}, and 8~\cite{HaywardPhasetransitionsBaTiO32005}.
The entries in the MPD that are taken from the ICSD database have gone through an energy relaxation step using density functional theory (DFT)~\cite{HohenbergInhomogeneousElectronGas1964,KohnSelfConsistentEquationsIncluding1965} before the crystal structures are deposited in the MPD. For some reason, the DFT relaxation took some of the lattice parameters somewhat far away from the experimental values in the original structure reports~\cite{XiaoCrystalstructuredense2008,Donohueeffectvarioussubstituents1958,HaywardPhasetransitionsBaTiO32005}.~
Overall the heuristic-1 approach already returned the correct structures for BaTiO$_3$ nanoparticles. The complete mining operation took 29.3 seconds when searching with the MPD and 47.8 seconds for the COD search to complete, using a general laptop.

\begin{figure}
\includegraphics[width=1\textwidth]{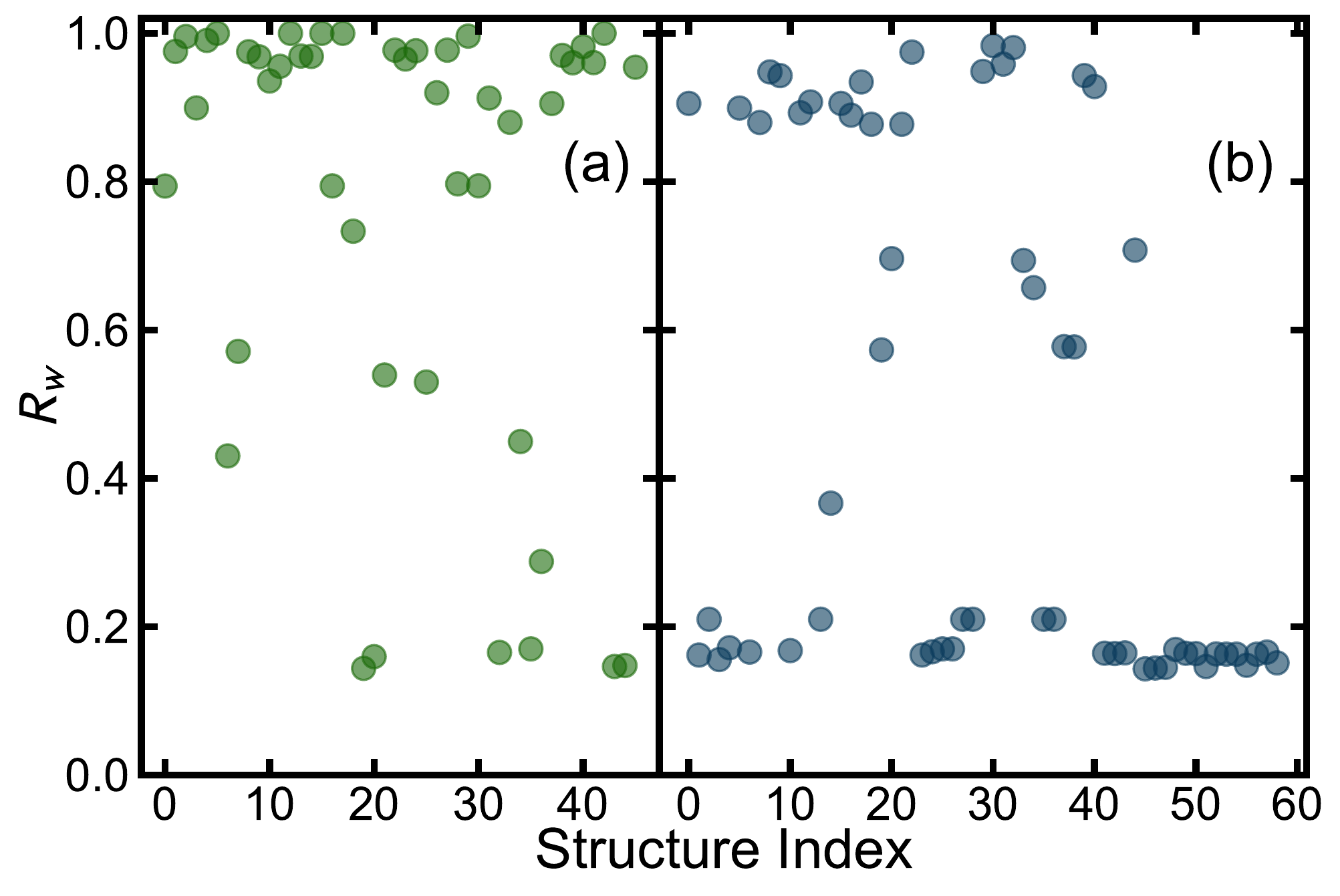}\\
\caption{\rw values for each of the structures \sout{pulled}\ins{fetched} from the databases for the BaTiO$_3$ nanoparticle x-ray data using heuristic-2, filtering for all the structures with Ba, Ti, and O elements from (a) the MPD (green) and (b) the COD (blue).
}
\label{fig;BTO_BaTiO_MPD_COD}
\end{figure}

We would like to further test the more loosely filtered heuristic-2 approach on the BaTiO$_3$ nanoparticle data. The \sm results from the MPD and COD, fetching all structures that contain just Ba, Ti, and O elements with any \sto, are shown in Fig.~\ref{fig;BTO_BaTiO_MPD_COD}(a) and (b), respectively. More details about the results can be found in the supporting information CSV files.

Heuristic-2 found all the structures that were found with heuristic-1, as expected.
This approach also found a number of additional good structural candidates.
The MPD returned three more that were within $\Delta R_w \approx 0.1$ from the best-fit \rw (approximately 0.14), i.e. MPD No.~43 (Ba$_{12}$Ti$_{12}$O$_{27}$)~\cite{JainCommentaryMaterialsProject2013a}, 44 (Ba$_{3}$Ti$_{3}$O$_{8}$)~\cite{WoodwardVacancyorderingreduced2004}, 36 (Ba$_{4}$Ti$_{4}$O$_{10}$)~\cite{JainCommentaryMaterialsProject2013a} and COD returned one, No.~4 model (Ba$_{0.92}$Ti$_{0.9}$O$_{2.89}$)~\cite{WadaCrystalstructurebarium2000},  where $\Delta R_w$ is the deviation in $R_w$ of a structure from the $R_w$ of the best-fit structure.
\ins{Close inspection of these models indicates that they have a stoichiometry that is approximately the Ba:Ti:O = 1:1:3 ratio and that \sm found some nearby defective structures in addition to the standard 113 perovskite structures. This will allow the experimenters to further investigate the defective models to find any physical or chemical insights that they might provide.}

The heuristic-2 structure-mining operation also, as expected, returned some structures from the databases for which the atomic composition ratio was not close to 1:1:3. None of these additional structures gave reasonable fits to the PDF, resulting in poor \rw values larger than 0.4 for the MPD (such as Ba$_{2}$Ti$_{3}$O$_{8}$ MPD No.~6~\cite{JainCommentaryMaterialsProject2013a}) and 0.6 for the COD (such as Ba$_{11}$Ti$_{28}$O$_{66.48}$ COD No.~34~\cite{VanderahCrystalStructureProperties2004}).
The entire search process took 493.7 seconds for the MPD and 469.5 seconds for the COD.

The heuristic-3 approach was also tested on the BaTiO$_3$ nanoparticle data by \sout{pulling}\ins{fetching} all structures that contain Ba, Ti, O elements and one additional element with any \sto. More details about the results can be found in the supporting information CSV files. It took about 10.3 and 41.0 minutes for the MPD (\sout{pulled totally}\ins{in total} 57 structures) and COD (\sout{pulled totally}\ins{in total} 103 structures) to finish, respectively. Of these new structures that were found, most of the best-fit structures have slightly worse $R_w$ ($\sim 0.2$) than those in heuristic-1 and 2 ($\sim 0.14$).  The new structures \sout{pulled}are mostly substituting Ba or Ti site by another element and they also have an approximate \sto 113, such as MPD No.~43 (Ba$_{3}$Sr$_{5}$Ti$_8$O$_{24}$)~\cite{JainCommentaryMaterialsProject2013a} and COD No.~22 (Ba$_{0.93}$ Ti$_{0.79}$ Mg$_{0.21}$ O$_{2.97}$)~\cite{WadaCrystalstructurebarium2000}, which agrees with what has been found in heuristic-2.

\begin{figure}
\includegraphics[width=1\textwidth]{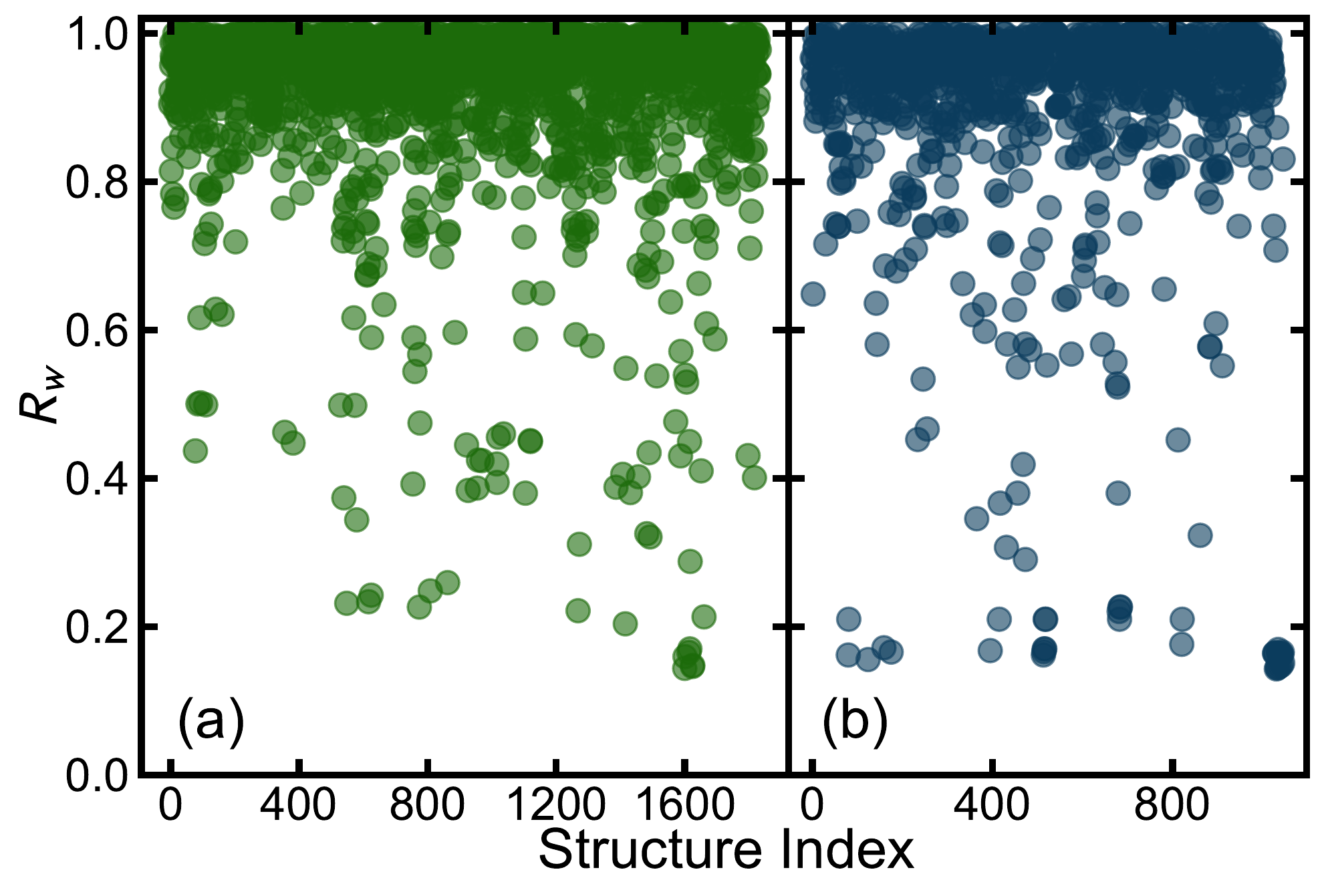}\\
\caption{\rw values for each of the structures \sout{pulled}\ins{fetched} from the databases for the BaTiO$_3$ nanoparticle x-ray data using heuristic-4, \sout{fetching}\ins{filtering for} all the structures with Ba, and two other arbitrary elements from (a) the MPD (green) and (b) the COD (blue).
}
\label{fig;BTO_BaXX_MPD_COD}
\end{figure}

Finally we tested the very loose heuristic-4 approach. Here the experimenter can freely choose any searching criteria, such as Ba-Ti-*, Ba-*-O, or even *-*-*, in which \ins{an} * represents an arbitrary element.
In our \ins{test} case we set the search to be that where the structure contains \sout{Ba and two other arbitrary elements with any \sto}\ins{three elements, including Ba and two other elements}, i.e. Ba-*-*. The \sm map plot is shown in Fig.~\ref{fig;BTO_BaXX_MPD_COD}.
This search took much longer, 174.3 and 205.2 minutes on a single CPU core for the MPD and COD, respectively. This may be sped up by running on more cores. \sout{Totally}\ins{In total,} 1833 structures were \sout{pulled}\ins{fetched} from the MPD and 1046 from the COD.  More details about the results are available in the supporting information CSV files.
The less restrictive heuristic-4 found all the structures that were found with heuristic-1 and 2, as expected.
The normal BaTiO$_3$ perovskite structures are still ranked at the top. Following that, it additionally returns some perovskite structures that have Ti replaced by other species with similar x-ray scattering power as Ti, such as MPD No.~1660 (BaVO$_3$)~\cite{NishimuraHighpressuresynthesisBaVO32014}, MPD No.~1268 (BaMnO$_3$)~\cite{JainCommentaryMaterialsProject2013a}, and COD No.~683 (BaFeO$_3$)~\cite{ErchakReactionFerricOxide1946b}. These gave agreements of $R_w \gtrsim 0.2$ compared to 0.14 for the best-fit structures (BaTiO$_3$). So the \sm is able to distinguish these nearby but incorrect structures from the ones with correct atom species.
The perovskite structures with B site element replaced by one with a significantly different x-ray scattering power than Ti resulted in significantly poorer \rw, away from the best-fit structures by $\Delta R_w \sim 0.15$, such as MPD No.~1482 (BaRhO$_3$)~\cite{BalachandranDefectGenomeCubic2017a} and COD No.~431 (BaNbO$_3$)~\cite{GrinBaNb3O6istPerowskit2014a}.

Overall we achieved a satisfactory result for the barium titanate nanoparticle dataset using all the four \sm heuristics.

\begin{figure}
\includegraphics[width=1\textwidth]{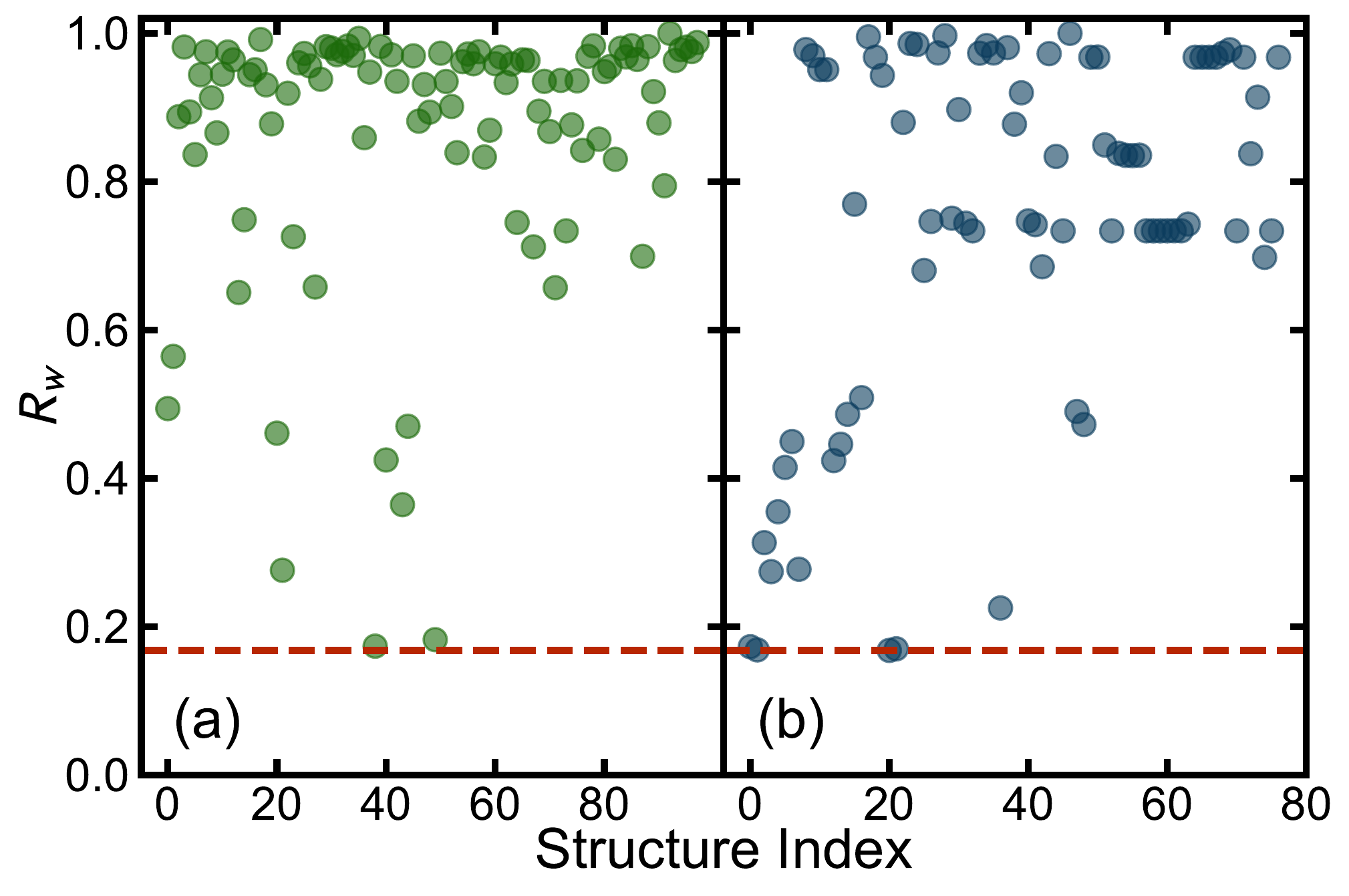}\\
\caption{\rw values for each of the structures \sout{pulled}\ins{fetched} from the databases for the Ti$_4$O$_7$ x-ray data using heuristic-2, \sout{fetching}\ins{filtering for} all the structures with Ti and O elements from (a) the MPD (green) and (b) the COD (blue). The horizontal dashed line represents the lowest $R_w$ entry found, COD No.~20 (Ti$_4$O$_7$, s.g.: \textit{P}$\overline{1}$).
}
\label{fig;Ti4O7_TiO_MPD_COD}
\end{figure}

We now test \sm for some different \sout{structures}\ins{materials}, for example, the low symmetry Ti$_4$O$_7$ system. Its published room temperature crystal structure is a triclinic model (space group \textit{P}$\overline{1}$) with all the atoms sitting on ($x$,$y$,$z$) general positions~\cite{MareziocrystalstructureTi4O71971a}. We used the \sm heuristic-2 approach, \sout{pulling}\ins{fetching} all the structures that contain Ti and O elements with any \sto.
The \sm map plot is shown in Fig.~\ref{fig;Ti4O7_TiO_MPD_COD} and the detailed results are available in the supporting information CSV files.
The top seven \sm results are also summarized in Table~\ref{tab;Ti4O7_top7}.
\begin{table*}
\small
\centering
\floatcaption{The top seven \sm results for the Ti$_4$O$_7$ experimental x-ray PDF using heuristic-2 on data from the MPD and COD. See the caption of Table~\ref{tab;BTO_BaTiO3_MPD} for an explanation of the entries. The full table can be found in the supporting information CSV files. The initial lattice parameters and refined ADPs are listed. The refined lattice parameters are not listed because they are close to initial values.}
\label{tab;Ti4O7_top7}
\scalebox{0.8}{
\begin{tabular}{ccccccccccccc}
\toprule
{DB No.} &        \rw & formula & s.g. &    Ti $U_{iso}$ &   O $U_{iso}$ &       $a_i$ &       $b_i$ &        $c_i$ &    $\alpha_i$ &     $\beta_i$ &     $\gamma_i$ & Ref. \\
 &  & & & {(\AA$^{2}$)} & {(\AA$^{2}$)} & {(\AA)}  & {(\AA)} & {(\AA)} & {($^{\circ}$)} & {($^{\circ}$)} & {($^{\circ}$)} & \\
\midrule
COD 20 &  0.168 &   Ti$_{4}$O$_{7}$ &  $P\overline{1}$ &  0.0051 &   0.0076 &  5.60 &  7.13 &  12.47 &  95.1 &  95.2 &  108.7 & \cite{MareziocrystalstructureTi4O71971a}\\
COD 1 &  0.169 &   Ti$_{4}$O$_{7}$ &  $P\overline{1}$  &  0.0050 &   0.0104 &  5.59 &  6.91 &   7.13 &  64.1 &  71.0 &   75.3 & \cite{HodeauStructuralaspectsmetalinsulator1979}\\
COD 21 &  0.170 &   Ti$_{4}$O$_{7}$ &  $P\overline{1}$ &  0.0050 &   0.0104 &  5.59 &  6.91 &   7.13 &  64.1 &  71.1 &   75.5 & \cite{MarezioStructuralaspectsmetalinsulator1973}\\
COD 0 &  0.173 &   Ti$_{4}$O$_{7}$ &  $P\overline{1}$  &  0.0048 &   0.0108 &  5.59 &  6.90 &   7.12 &  64.1 &  71.2 &   75.7 & \cite{HodeauStructuralaspectsmetalinsulator1979}\\
MPD 38 &  0.174 &   Ti$_{5}$O$_{9}$ &  $P\overline{1}$ &  0.0046 &   0.0065 &  5.62 &  7.18 &   8.56 &  69.5 &  75.2 &   71.3 & \cite{MarezioPhasetransitionsmathrmTi1977}\\
MPD 49 &  0.183 &   Ti$_{4}$O$_{7}$ &  $P\overline{1}$ &  0.0048 &   0.0108 &  5.64 &  6.96 &   7.18 &  64.2 &  71.1 &   75.1 & \cite{HodeauStructuralaspectsmetalinsulator1979}\\
COD 36 &  0.225 &   Ti$_{5}$O$_{9}$ &   $P1$           &  0.0053 &   0.0088 &  5.57 &  7.12 &   8.49 &  69.8 &  75.0 &   71.5 & \cite{AnderssonThecrystalstructureofti5o91960}\\
\vdots & & & & & & & & & & & & \\
\bottomrule
\end{tabular}
}
\end{table*}
The titanium oxides have many different structures, largely depending on the stoichiometry (98 structures  from the MPD and 77 from the COD), but \sm returned the published structure for Ti$_{4}$O$_{7}$ on the top, i.e. COD No.~20~\cite{MareziocrystalstructureTi4O71971a}.

This is a challenging problem because there are similar structures belonging to the Ti$_n$O$_{2n-1}$ Magn\'{e}li homologous series~\cite{AnderssonDiskreteTitanoxydphasenim1956,andersson1957phase}.
Among the top 7 entries, the other 4 Ti$_4$O$_7$ structures are very similar to COD No.~20.  COD~20 is reported in a different structural setting than the other 4~\cite{SetyawanHighthroughputelectronicband2010a}, which explains the rather different values for the lattice parameters, but the only real difference in structure between COD~20 and the other Ti$_{4}$O$_{7}$ structures reported in Table~\ref{tab;Ti4O7_top7} is that one oxygen position is shifted by about 0.7~\AA\ along the $b$-axis compared to the other four. This is a significant structural difference yet does not result in a very large difference in $R_w$ and so differentiating these two structures probably deserves some additional consideration by the experimenter. Atomic positions are not refined independently during \sout{the}\ins{this} \sm process and it is possible that this discrepancy may be resolved by a full refinement of the best performing models, as well as suggesting to the user oxygen $b$-axis position as a possibly relevant variable.
Structure-mining also returned some results with slightly different stoichiometry with similar $R_w$ values.  For example, the MPD No.~38 (Ti$_{5}$O$_{9}$)~\cite{MarezioPhasetransitionsmathrmTi1977}, which belongs to a different variant in the Magn\'{e}li series.  The Magn\'{e}li phases are constructed from similar TiO$_6$ octahedral motifs, containing rutile-like slabs extending infinitely in the $a$-$b$ plane, but the TiO$_6$ octahedra are stacked along the $c$-axis
in slabs of different widths depending on the composition~\cite{AnderssonDiskreteTitanoxydphasenim1956,andersson1957phase,MarezioPhasetransitionsmathrmTi1977}. In Ti$_{4}$O$_{7}$, every oxygen atom connects four octahedra, but in Ti$_5$O$_9$ (MPD~38), oxygen atoms link 3 octahedra.  Despite these differences, the MPD~38 model performs similarly, albeit somewhat worse, than some of the well performing Ti$_{4}$O$_{7}$ models, suggesting that it at least warrants being explicitly ruled out as a candidate in a more careful modeling.
This illustrates how the \sm approach, beyond just automatically finding the ``right" structure, additionally can add value by suggesting alternative nearby models to the experimenter.
We also note that, from Table~\ref{tab;Ti4O7_top7}, COD No.~36 (Ti$_{5}$O$_{9}$, s.g.: \textit{P}1)~\cite{AnderssonThecrystalstructureofti5o91960} performs worse ($R_w > 0.2$), and it is the first model that has a significantly different structure, where some Ti atoms are tetrahedrally coordinated by oxygen rather than octahedrally.  This model can probably be ruled out on the basis of \sm alone.

Now let us turn to a challenging dataset, nanowire bundles of a pyroxene compound with a generic composition of XYSi$_2$O$_6$ (where X and Y refer to metallic elements such as but not limited to Co, Na, and Fe).
This example is particularly challenging because the samples formed as nanowires that were reported to be $\sim 3$~nm in width~\cite{lewis;cec18}. In that work, a series of candidate structures were tried manually and the best-fit model was found to be monoclinic NaFeSi$_2$O$_6$ with a space group \textit{C}2/\textit{c}~\cite{pap1969crystal}.

The \sm heuristic-1 approach was first tested. The MPD found one structure~\cite{pap1969crystal} and the COD found six non-duplicated structures~\cite{Suenohightemperaturecrystal1973,thompson2004model,RedhammerSynthesisstructuralproperties2000a,redhammer2006single,nestola2007crystal,mccarthy2008situ}, all having a quite similar structure, NaFeSi$_2$O$_6$ (s.g.: \textit{C}2/\textit{c}). The returned \sm results have $R_w \approx 0.35$. These are poor fits overall, but comparable to the fits reported in the prior work~\cite{lewis;cec18}. Although the \rw is not ideal, possibly due to the sample's complicated geometry, structural heterogeneity, and defects, the \sm approach seems still to be working.
Using heuristic-2 (Na-Fe-Si-O) and 3 (Na-Fe-Si-O-*) approaches found similar results, with heuristic-3 finding some Ca and Li doped compounds albeit with the same structure.

\begin{figure}
\includegraphics[width=1\textwidth]{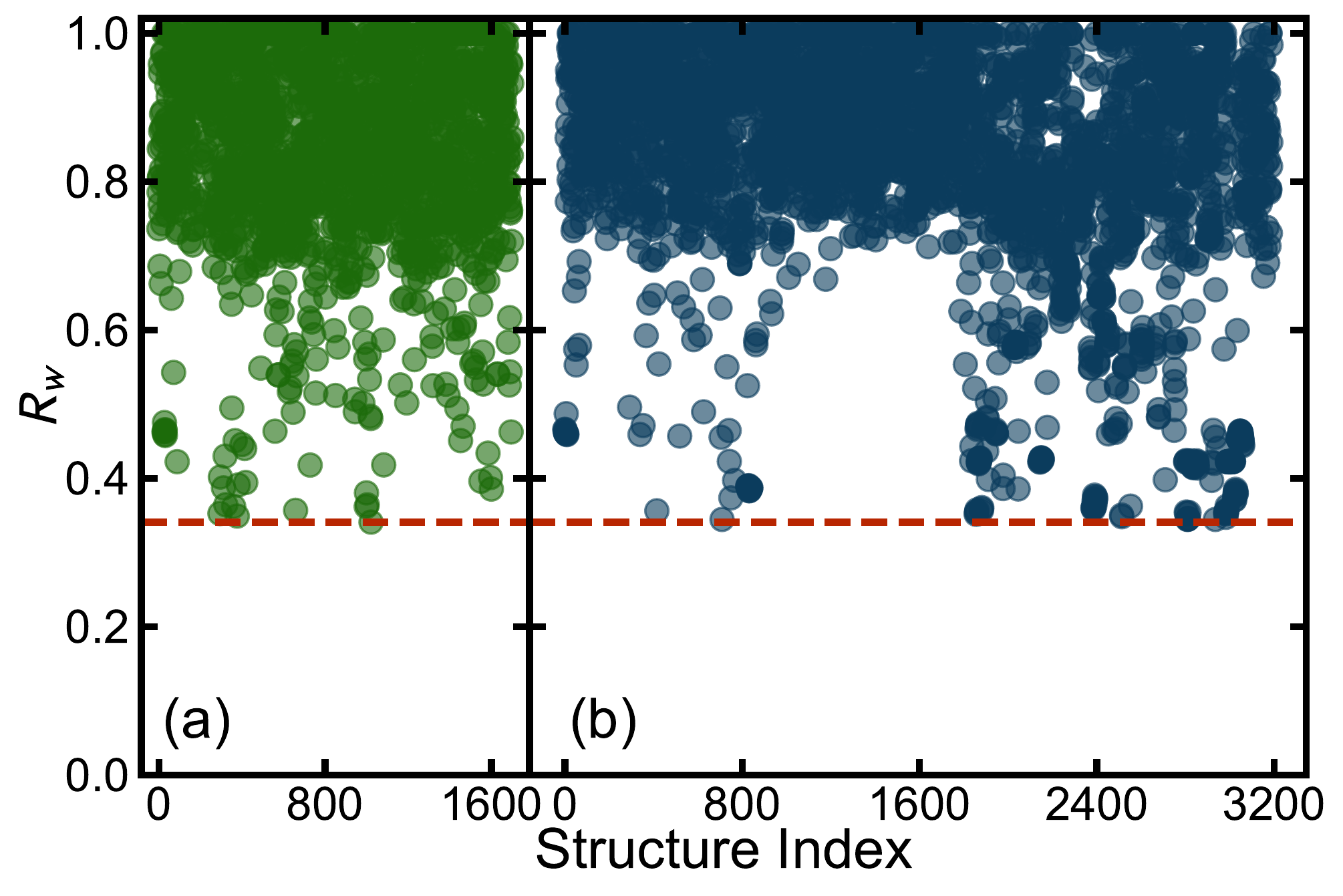}\\
\caption{\rw values for each of the structures \sout{pulled}\ins{fetched} from the databases for the NaFeSi$_2$O$_6$ nanowire x-ray data using heuristic-4, \sout{fetching}\ins{filtering for} all the structures with Si, O, and two other arbitrary elements from (a) the MPD (green) and (b) the COD (blue). The horizontal dashed line represents the lowest $R_w$ entry found, MPD No.~1021 (NaGaSi$_2$O$_{6}$, s.g.: \textit{C}2/\textit{c}).
}
\label{fig;nanowire_XXSiO_MPD_COD}
\end{figure}

The least restrictive heuristic-4 approach was also tried. Here we show the result of fetching all the structures that contain Si and O elements and two other arbitrary elements with any \sto, i.e. *-*-Si-O (Fig.~\ref{fig;nanowire_XXSiO_MPD_COD}).
The mining operation took about 12 hours for the MPD (\sout{pulled}in total 1700 structures) and 122 hours for the COD (\sout{in total}3187 structures) to finish, respectively. The COD is significantly more time-consuming because many of the COD \sout{pulled}structures have large numbers of hydrogen atoms, which could be neglected for x-ray PDF calculation to shorten the running time in future work. More details about the results are available in the supporting information CSV files. However, the top ten entries across the MPD and COD are listed here for convenience in Table~\ref{tab;nanowire_XXSiO_top10}.
%
\begin{table*}
\small
\centering
\floatcaption{The top ten \sm results for the NaFeSi$_2$O$_6$ nanowire experimental x-ray PDF using heuristic-4 on data from the MPD and COD, \sout{pulling}\ins{fetching} all the structures that contain Si and O elements and two other arbitrary elements with any \sto, i.e. *1-*2-Si-O. The *1 and *2 represent the first and the second atoms in the formula, respectively. See the caption of Table~\ref{tab;BTO_BaTiO3_MPD} for an explanation of the entries. The full table can be found in the supporting information CSV files. The refined lattice parameters and ADPs are listed. \sout{The initial lattice parameters are not listed because they are close to refined values and the refined lattice parameters are mostly slightly larger than the initial values.}\ins{The initial lattice parameters are available in the supporting information but are not reproduced here. They are all very close to refined values.}
}
\label{tab;nanowire_XXSiO_top10}
\scalebox{0.8}{
\begin{tabular}{cccccccccccccc}
\toprule
{DB No.} &     \rw &       formula &  s.g. &  *1 $U_{iso}$ &  *2 $U_{iso}$ &  Si $U_{iso}$ &  O $U_{iso}$ &   $a_r$ &   $b_r$ &   $c_r$ &  $\beta_r$ &   SPD &      DB ID \\
 &  & & & {(\AA$^{2}$)} & {(\AA$^{2}$)} & {(\AA$^{2}$)} & {(\AA$^{2}$)} & {(\AA)} & {(\AA)} & {(\AA)} & {($^{\circ}$)} & {(\AA)} & \\
\midrule
 MPD 1021 &  0.341 &      NaGaSi$_2$O$_6$     &  $C2/c$ &    0.0193 &    0.0053 &    0.0048 &  0.0118 &    9.69 &  8.81 &  5.32 &     107.5 &  34.0 &    mp-6822 \\
  COD 709 &  0.345 &     NaGaSi$_2$O$_6$      &  $C2/c$ &    0.0174 &    0.0054 &    0.0049 &  0.0112 &    9.68 &  8.81 &  5.32 &     107.5 &  33.8 &    2004306 \\
 COD 2935 &  0.345 &     NaGaSi$_2$O$_6$      &  $C2/c$ &    0.0174 &    0.0054 &    0.0049 &  0.0112 &    9.68 &  8.81 &  5.32 &     107.5 &  33.8 &    9011383 \\
 COD 2809 &  0.345 &      NaGaSi$_2$O$_6$     &  $C2/c$ &    0.0173 &    0.0054 &    0.0048 &  0.0112 &    9.68 &  8.81 &  5.32 &     107.5 &  33.8 &    9010186 \\
 COD 2983 &  0.348 &     NaFeSi$_2$O$_6$      &  $C2/c$ &    0.0249 &    0.0033 &    0.0088 &  0.0129 &    9.68 &  8.82 &  5.32 &     107.5 &  34.1 &    9013274 \\
 COD 2513 &  0.348 &     NaFeSi$_2$O$_6$      &  $C2/c$ &    0.0214 &    0.0035 &    0.0070 &  0.0144 &    9.68 &  8.82 &  5.32 &     107.5 &  34.7 &    9005439 \\
  MPD 377 &  0.349 &  Ca$_{0.5}$NiSi$_2$O$_6$ &    $C2$ &    0.0118 &    0.0041 &    0.0052 &  0.0136 &    9.68 &  8.81 &  5.31 &     107.4 &  32.9 &  mvc-12761 \\
 COD 1856 &  0.352 &     NaFeSi$_2$O$_6$      &  $C2/c$ &    0.0221 &    0.0033 &    0.0079 &  0.0137 &    9.69 &  8.81 &  5.32 &     107.6 &  34.6 &    9000327 \\
 COD 2805 &  0.353 &      NaFeSi$_2$O$_6$     &  $C2/c$ &    0.0227 &    0.0032 &    0.0082 &  0.0135 &    9.69 &  8.81 &  5.32 &     107.6 &  34.7 &    9010095 \\
  MPD 294 &  0.353 &  Ca$_{0.5}$CoSi$_2$O$_6$ &    $C2$ &    0.0277 &    0.0042 &    0.0050 &  0.0231 &    9.68 &  8.82 &  5.32 &     107.3 &  34.8 &  mvc-11818 \\
\vdots &&&&&&&&&&&&& \\
\bottomrule
\end{tabular}
}
\end{table*}

The returned NaGaSi$_2$O$_6$ entries (s.g.:\textit{C}2/\textit{c})~\cite{OhashicrystalstructureNaGaSi2O61983,OhashiLowDensityFormNaGaSi2O61995,NestolaLowtemperaturebehaviorNaGaSi2O62007a} have a similar structure to NaFeSi$_2$O$_6$ (s.g.:\textit{C}2/\textit{c}). They both fit experimental data comparably well with NaGaSi$_2$O$_6$ slightly preferred. The NaGaSi$_2$O$_6$ solution can be ruled out on the basis that no Ga was in the synthesis. The x-ray scattering power of Fe and Ga are similar with Ga being slightly higher ($Z(Fe) =26$, $Z(Ga) = 31$). The fact that \sm prefers to put a slightly higher atomic number, $Z$, element at this position suggests that we have the right structure, but some details of the refinement need to be worked out by the experimenter.
\sout{Structure-mining also suggests that the refined lattice parameters are mostly slightly larger than the initial values.}This example illustrates how careful interrogation of the fits to the \sout{pulled}database models compared to the original parameters can highlight possible defects or impurities and guide the experimenter towards what things to search for.

The MPD also returned some computed theoretical structures with space group \textit{C}2, MPD No.~377 (Ca$_{0.5}$NiSi$_{2}$O$_{6}$, s.g.: \textit{C}2) and MPD No.~294 (Ca$_{0.5}$CoSi$_{2}$O$_{6}$, s.g.: \textit{C}2)~\cite{JainCommentaryMaterialsProject2013a}. These perform slightly worse than the fully stoichiometric NaGaSi$_2$O$_6$ and NaFeSi$_2$O$_6$ structures.  Inspection of these structures indicates that they are very similar in nature but with a lowered symmetry due to missing Ca ions and can probably be ruled out, though the fact that \sm finds them may suggest trying sub-stoichiometry models on the \sout{alkali metal site}\ins{A site}.

Overall, the heuristic-4 returned a number of isostructural but with different composition structures. For this system, it is possible that the \sout{ground truth answer}\ins{correct structure} is not limited to the pure NaFeSi$_2$O$_6$ (s.g.: \textit{C}2/c) stoichiometry only and substituting impurity ions or atom deficiencies may be occuring for such a complicated synthesis~\cite{lewis;cec18}. These candidate structures found by \sm are valuable to resolve the ambiguity. Furthermore, by taking the \sm approach yields different but similarly-fitting models which can also give meaningful information about uncertainty estimates on refined parameters such as metal or oxygen ion positions.
This test again shows the huge potential of \sm on PDF data to help experimenters be aware of some possible structural solutions that were overlooked or not realized in the traditional workflow.

\begin{figure}
\includegraphics[width=1\textwidth]{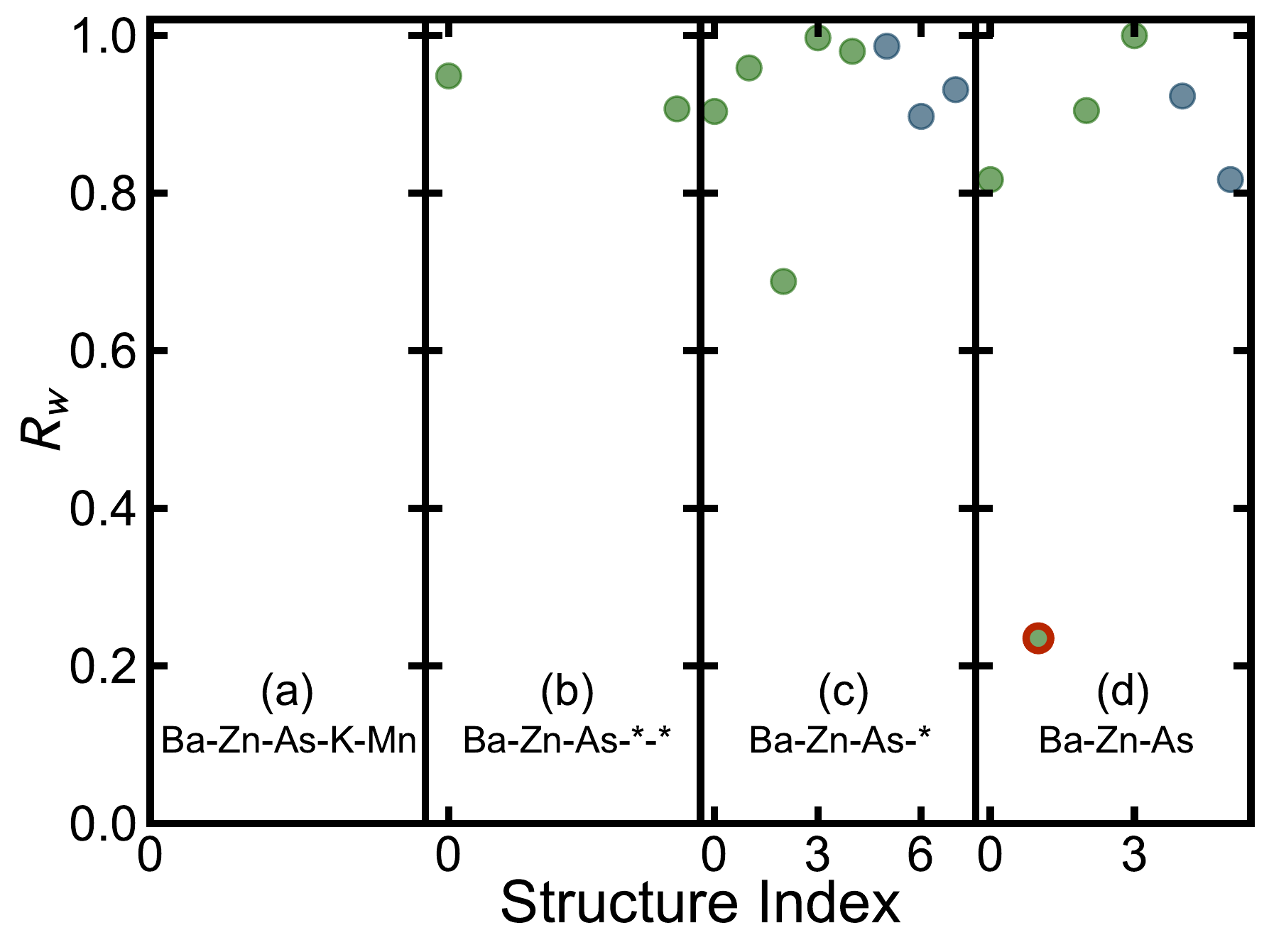}\\
\caption{\rw values for each of the structures \sout{pulled}\ins{fetched} from the databases for the Ba$_{0.8}$K$_{0.2}$(Zn$_{0.85}$Mn$_{0.15}$)$_2$As$_2$ neutron data \sout{fetching}\ins{using the heuristics of} (a) Ba-Zn-As-K-Mn, (b) Ba-Zn-As-*-*, (c) Ba-Zn-As-*, and (d) Ba-Zn-As from the MPD (green) and the COD (blue). The best-fit model MPD No.~1 (BaZn$_2$As$_2$) in (d) is marked by a red circle.
}
\label{fig;Ba122K02Mn015_MPD_COD}
\end{figure}

Next, we test \sm on a complicated doped material, Ba$_{1-x}$K$_{x}$(Zn$_{1-y}$Mn$_y$)$_2$As$_2$.
We used the neutron PDF data with composition $(x,y) = (0.2,0.15)$, which has both A-site and B-site dopings.
Its published room temperature crystal structure is a tetragonal structure with the space group \textit{I}4/\textit{mmm}~\cite{frand;prb16}.
First we applied heuristic-2 specifying all the elements including the dopants, i.e. fetching Ba-Zn-As-K-Mn structures regradless of \sto.
This returned no structures from the MPD or the COD.
We next tested a heuristic-4 approach with Ba-Zn-As-*-*.  This did result in two structures being returned, but they were both incorrect compounds, Ba$_2$MnZn$_2$As$_2$O$_2$~\cite{OzawaSynthesisCharacterizationNew1998} and BaZn$_2$As$_3$HO$_{11}$~\cite{JainCommentaryMaterialsProject2013a}, with $R_w$ values close to 1, as shown in Fig.~\ref{fig;Ba122K02Mn015_MPD_COD}(b).
\sout{Additionally the heuristic-4 approach was tested to look for a sample with doping on only one site (Ba-Zn-As-*), but still found only incorrect structures, as shown in Fig.~\ref{fig;Ba122K02Mn015_MPD_COD}(c).}
\ins{We then looked for structures with doping on one site. The ``Ba-Zn-As-*" searches the databases for compositions containing four elements, including Ba, Zn, As and one other element. But it still only found incorrect structures, as shown in Fig.~\ref{fig;Ba122K02Mn015_MPD_COD}(c).}
Finally, we resorted to a heuristic-2 approach but only giving the composition of the undoped endmember, Ba-Zn-As.  This did find the correct structure, tetragonal phase MPD No.~1 (BaZn$_2$As$_2$, s.g.: \textit{I}4/\textit{mmm})~\cite{hellmann2007neue}, as marked by the red circle in Fig.~\ref{fig;Ba122K02Mn015_MPD_COD}(d),
even though we were fitting to the doped data.
This suggests a feasible strategy for doped systems if they are not represented in the databases, which is to try searching for the parent undoped structure, on the basis that the doped structure may be still close to its parent phase, regardless of possible local structure distortions introduced by doping~\cite{frand;prb16}.
Starting from this success, the experimenter could then easily change the occupancy of the A-site or B-site, which was also how \sout{people performed structural analysis}\ins{structural analysis was previously performed} on this doped material~\cite{ZhaoNewdilutedferromagnetic2013,RotterSuperconductivity38Iron2008}.
\sout{So \sm has been proved to work well even for the complicated doped system.}\ins{So even for the case of doped structures, \sm found the correct geometric structure which was from the nearest undoped variant in the database (in this case, there were no structures in the databases that had the same composition as the measured sample). The experimenter can take this structure model and introduce dopants with the known composition.}

\ins{
Note that the PDF measures the {\it local} structure of materials, which is not necessarily identical to the long-range ordered crystal structure in some materials~\cite{Keencrystallographycorrelateddisorder2015a,BozinLocalorbitaldegeneracy2019c}.
As a result, \sm will search for the closest structure in the crystal structure databases to the local atomic arrangement.
Here we present the case of the CuIr$_2$S$_4$ system, which has a tetragonal local structure (s.g.: $I4_1/amd$) but a cubic long-range ordering (s.g.: $Fd\overline{3}m$) above the metal-insulator transition temperature ($T_{MIT} = 226$~K)~\cite{BozinLocalorbitaldegeneracy2019c}.
The x-ray data measured at 500~K was tested. when fitting over the narrow range of $1.5<r<5$~\AA, $1.5<r<10$~\AA, or $1.5<r<20$~\AA, \sm ranks the tetragonal model above the cubic model.
However, when fitting over the broad range of $1.5<r<50$~\AA, the cubic model fits slightly better than the tetragonal model.
The representative results for the $1.5<r<5$~\AA\ and $1.5<r<50$~\AA\ fit ranges can be found in the supporting information CSV files.
As a result, \sm returns the symmetry broken and non-symmetry broken structural candidates whether it is fit over a narrow or broad range, but it also ranks them correctly depending on the $r$-range fit over in this test case. We note that it is possible in \sm for a user to specify a custom fit range, which would allow the researcher to search for structures that are relevant for the measured PDF on different length-scales.
}

\sout{Finally, we would like to}\ins{We would also like to} test the robustness of the \sm approach when the structural data also include non-structural signals, such as the magnetic PDF (mPDF) signal~\cite{frand;aca14,frand;aca15,frand;prl16} in a neutron diffraction experiment of a magnetic material.
To test this we consider the MnO neutron PDF data, measured at 15~K, which has a strong mPDF signal.
Early neutron diffraction studies reported that MnO has a cubic structure in space group \textit{F}\textit{m}$\overline{3}$\textit{m} at high temperature and undergoes an antiferromagnetic transition with a N\'{e}el temperature of $T_{N} = 118$~K, which results in a rhombohedral structure in space group \textit{R}$\overline{3}$\textit{m}~\cite{ShullNeutronDiffractionParamagnetic1951,RothMagneticStructuresMnO1958}. More recently it has been suggested that, at low-temperature, the local structure is even lower symmetry, e.g., monoclinic in s.g. \textit{C}2~\cite{GoodwinMagneticStructureMnO2006,frand;aca15}.  Here we see which of these structural results are returned by the \sm process.

\begin{figure}
\includegraphics[width=1\textwidth]{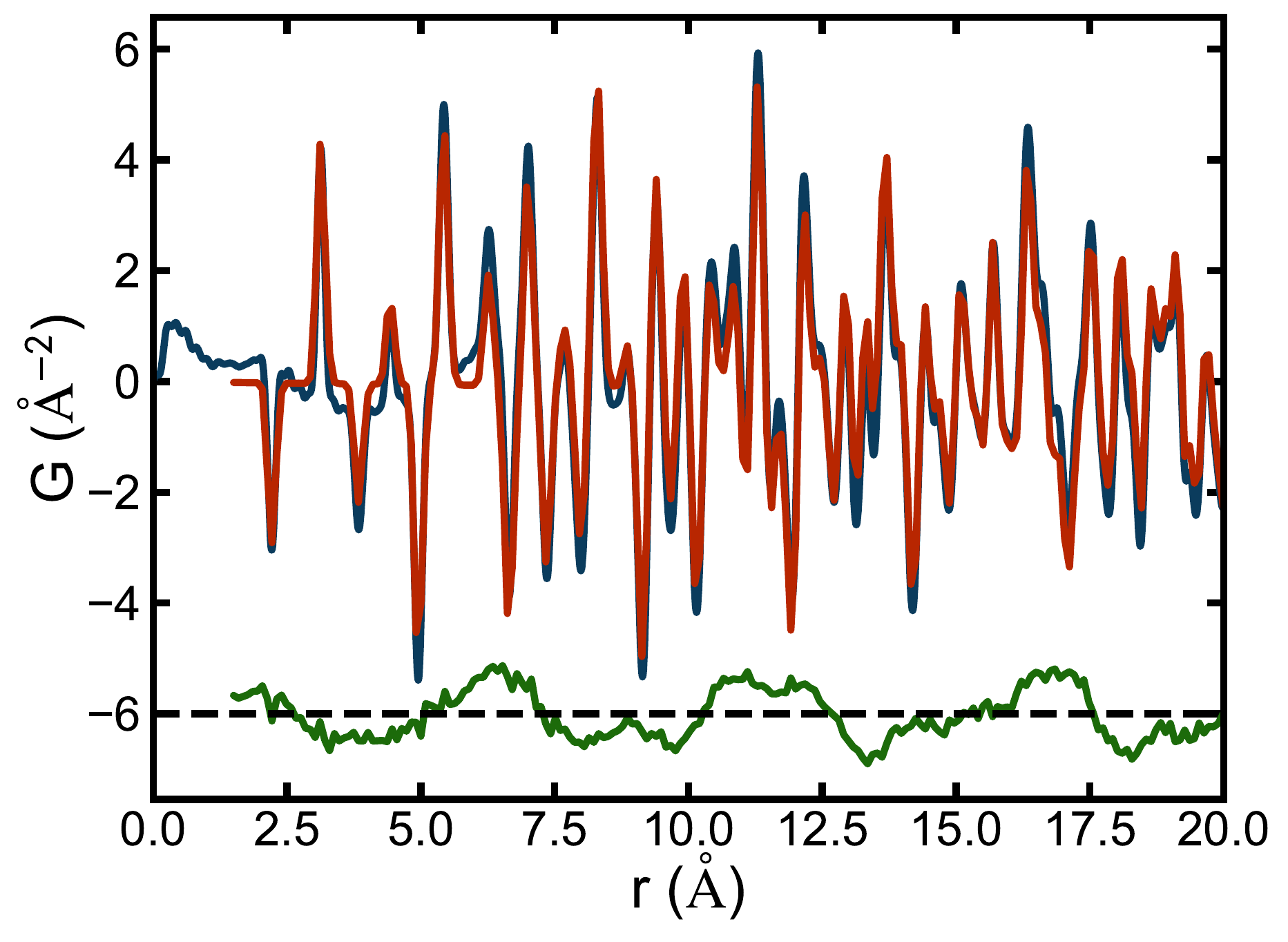}\\
\caption{The neutron PDF of the MnO data (blue curve) measured at 15~K with the best-fit calculated atomic PDF (red) for the MPD No.~41, rhombohedral MnO model from heuristic-2. The difference curve is shown offset below (green). Notice the strong magnetic PDF signal in the difference curve, which did not confuse \sm.
}
\label{fig;MnO_15K_fit}
\end{figure}

The heuristic-2 approach is applied, i.e. fetching all the atomic structures with Mn and O elements. The rhombohedral MnO model is the best performing model (MPD No.~41~\cite{JainCommentaryMaterialsProject2013a} with $R_w = 0.236$, Fig.~\ref{fig;MnO_15K_fit}).
The second best fit is the cubic MnO model (COD No.~56~\cite{ZhangRoomtemperaturecompressibilitiesMnO1999} with $R_w = 0.310$).  This correctly reflects the fact that at 15~K the material is expected to be in the rhombohedral phase. The monoclinic s.g. \textit{C}2 model was not returned by \sm but this is because it is not in any of the databases. The fit agreements are similar to those reported in \cite{frand;aca15} when the magnetic model is not included in the fit (as is the case here).  Therefore, even in the presence of significant magnetic scattering, \sm is able to find the correct solution.
\sout{Interestingly, the cubic model was not present in the MPD and the rhombohedral model was not present in the COD, and the full picture was only obtained by mining multiple databases.}

\ins{Structure-mining was conceived as a structure selection approach and not for finding multiple phases in a sample.  However, it is interesting to establish how well it performs when the PDF signal consists of more than one phase. For this test we use the x-ray PDF dataset of a vanadium nitride sample~\cite{Urbankowski2Dmolybdenumvanadium2017d}. In the original publication~\cite{Urbankowski2Dmolybdenumvanadium2017d} it had manually been assigned as consisting of a majority (64\%) of V$_2$N with a structure in space group \textit{P}$\overline{3}$1\textit{m}, and a minority (36\%) of VN (s.g.: \textit{F}\textit{m}$\overline{3}$\textit{m}).
First we applied the heuristic-2 procedure searching for V-N structures on the measured data. The \sm found the correct V$_2$N structure successfully with $R_w=0.29$ and with other structures being $R_w>0.7$.  The procedure did not find VN as a candidate structure.
We then subtracted the calculated V$_2$N structure from the measured PDF and carried out \sm on the difference.  All of the returned structures resulted in values of $R_w$ that were large (0.66 and higher) which is presumably because of the low signal to noise ratio in the subtracted data, and the fact that our definition of $R_w$ (Eq.~\ref{eq:GoodnessOfFit}) does not account for measurement noise.  Nonetheless, the top best-fit structure returned by \sm was exactly the correct cubic VN phase ($R_w=0.66$). More details about the results can be found in the supporting information CSV files.
Structure-mining was, therefore, successful at finding both the majority phase and the secondary phase. This shows that, at least in favorable circumstances, multi-phase samples may be successfully structure-mined.
}

\ins{
We note that it should be straightforward to extend the \sm methodology to study the PDFs of organic materials. However, this is not done in the current version.
First, for organic material data, the PDF peaks are sharp at low-$r$ (intra-molecular range) and broad at high-$r$ (inter-molecular range). To handle this correctly, different ADPs should be applied for two seperate regions. Especially the sharp intra-molecular peaks cannot be sufficiently fit, which usually require some special treatments on parameters such as the correlated atomic motion parameter $\delta$, sratio, and ADPs~\cite{PrillModellingpairdistribution2015d,Prillsolutionrefinementorganic2016a}. This requires a separation of the molecule from its neighbors which is currently done manually and an automated approach needs to be developed for \sm to work.
Second, the current heuristics, searching by compositions and elements, are not suitable for organic materials. For example, searching ``C-H-O" would return too many candidates, about 16000 entries from COD and MPD databases. Development of new search heuristics, such as searching by organic molecule name is possible but needs some future work.
Third, there are limited organic material entries in the currently supported COD and MPD databases. Supporting some more comprehensive organic structural databases, such as the Cambridge Structural Database (CSD)~\cite{AllenCambridgeStructuralDatabase2002b}, is necessary for finding organic compounds.
}

\ins{
We have shown that \sm is able to find the desired structures from the mine on a range of test cases.  We now consider its robustness against factors that might prove problematic, specifically data collected at a different temperature to the data in the mine and data measured under a range of different experimental conditions.
Structure-mining seems to work well on data collected at different temperatures and so is robust against differences in lattice parameters and ADPs due to temperature effects. This assertion is supported by the BaTiO$_3$ example described above where \sm found all the barium titanate structural variants, which were measured at a range of temperatures from 15~K (COD No.~24 in Table~\ref{tab;BTO_BaTiO3_COD})~\cite{kwei;jpc93} to 1000~K (COD No.~14 in Table~\ref{tab;BTO_BaTiO3_COD})~\cite{EdwardsStructureBariumTitanate1951} when compared to the data measured at room temperature.
Structure-mining also performed well in tests where data were coming from a wide range of different instruments and measurement conditions. In these tests the $Q_{max}$ values for the data varied \sout{from}\ins{between} 18.6~\AA$^{-1}< Q_{max}<25.0$~\AA$^{-1}$, and the ranges of instrument resolution parameters are $0.038<Q_{damp}<0.058$~\AA$^{-1}$ and $0.0<Q_{broad}<0.048$~\AA$^{-1}$~\cite{YangConfirmationdisorderedstructure2013e,frand;prb16,QuinsonSpatiallyLocalizedSynthesis2018b}. All of them worked well in \sm which successfully found the correct structures regardless of the fact that different $Q_{max}$ and instrument resolution parameters were in effect. Thus the method should work in general for many other instruments even when $Q_{damp}$ and $Q_{broad}$ vary from one instrument to another.}

\section{Conclusion}

In this paper, we have demonstrated a new approach, called \sm,  for automated screening of large numbers of candidate structures to the atomic pair distribution function (PDF) data, by automatically fetching candidate structures from structural databases and automatically performing PDF structure refinements to obtain the best agreement between calculated PDFs of the structures and the measured PDF under study.
The approach has been successfully tested on the PDFs of a variety of challenging materials, including complex oxide nanoparticles and nanowires, low-symmetry structures, complicated doped, magnetic, locally distorted and mixed phase materials.
This approach could greatly speed up and extend the traditional structure searching workflow and enable the possibility of highly automated and high-throughput real-time PDF analysis experiments in the future.





\ack{\paragraph{Acknowledgements}
The authors thank Emil S. {Bo\v zin} and Cedomir Petrovic for sharing their unpublished Ti$_4$O$_7$ \ins{and published CuIr$_2$S$_4$} x-ray experimental data, and thank Benjamin A. Frandsen for sharing his published Ba$_{0.8}$K$_{0.2}$(Zn$_{0.85}$Mn$_{0.15}$)$_2$As$_2$ and MnO neutron experimental data.
Work in the S.J.L.B. group was supported by the U.S. National Science Foundation through grant DMREF-1534910.
L.Y. and M.G.T. acknowledge support from the ORNL Graduate Opportunity (GO) program, which was funded by the Neutron science directorate, with support from the Scientific User Facilities Division, Office of Basic Energy Science, U.S. Department of Energy (DOE).
P.J. was supported by the New York State BNL Big Data Science Capital Project under the U.S. DOE Contract No. DE-SC0012704.
M.W.T. gratefully acknowledges support from BASF.
X-ray PDF measurements were conducted on beamline 28-ID-2 of the National Synchrotron Light Source II, a U.S. DOE Office of Science User Facility operated for the DOE Office of Science by Brookhaven National Laboratory under Contract No. DE-SC0012704.
Use of the NOMAD beamline of the Spallation Neutron Source, Oak Ridge National Laboratory, was sponsored by the Scientific User Facilities Division, Office of Basic Energy Science, U.S. DOE.
Use of the NPDF beamline at LANSCE was funded by the Office of Basic Energy Science, U.S. DOE. Los Alamos National Laboratory is operated by Los Alamos National Security LLC under DOE Contract No. DE-AC52-06NA25396.
}


\bibliographystyle{iucr}
\bibliography{yang_structuremining}


\appendix
\begin{table*}[h]
\centering
\floatcaption{The experimental PDF datasets for testing the \sm approach with relevant parameters. Here $L$ is the sample-to-detector distance and $Q_{damp}$  and $Q_{broad}$ are standard fitting parameters for the PDF that come primarily from instrumental resolution effects. The instrument resolution parameters and $L$ of the CuIr$_2$S$_4$ data are not available.}
\label{tab;sisamples}
\begin{threeparttable}
\scalebox{0.8}{
\begin{tabular}{cccccccc}

Composition & Scatterer & Beamline & $Q_{damp}$  & $Q_{broad}$  & $Q_{max}$ & X-ray wavelength & $L$\\
 &  & & {(\AA$^{-1}$)} & {(\AA$^{-1}$)} & {(\AA$^{-1}$)} & {(\AA)} & {(mm)}\\\hline
BaTiO$_3$\tnote{a} & x-ray & XPD & 0.037 & 0.017 & 24.0 & 0.1867 & 202.8031\\
Ti$_4$O$_7$ & x-ray & XPD & 0.041 & 0.009 & 25.0 & 0.1866 & 202.9990\\
NaFeSi$_2$O$_6$\tnote{b} & x-ray & XPD & 0.035 & 0.016 & 22.0 & 0.18288 & 204.2825 \\
Ba$_{0.8}$K$_{0.2}$(Zn$_{0.85}$Mn$_{0.15}$)$_2$As$_2$\tnote{c} & neutron & NOMAD & 0.018 & 0.019 & 20.0 & - & - \\
CuIr$_2$S$_4$\tnote{d} & x-ray & XPD & - & - & 25.0 & 0.183 & - \\
MnO\tnote{e} & neutron & NPDF & 0.0198& 0.0195 & 35.0 & - & - \\
V$_2$N+VN\tnote{f} & x-ray & XPD & 0.0369 & 0.0131 & 25.0 & 0.1847 & 205.3939\\
\hline
\end{tabular}

}
\begin{tablenotes}
\item [a] \cite{lombardi;cm19}.
\item [b] \cite{lewis;cec18}.
\item [c] \cite{frand;prb16}.
\item [d] \cite{BozinLocalorbitaldegeneracy2019c}.
\item [e] \cite{frand;aca15}.
\item [f] \cite{Urbankowski2Dmolybdenumvanadium2017d}.
\end{tablenotes}
\end{threeparttable}
\end{table*}





\end{document}